\documentclass[fleqn]{article}
\usepackage{graphicx}
\usepackage{epsfig}
\usepackage{amsmath}
\usepackage{amsthm}
\usepackage{amssymb}
\usepackage{enumerate}
\usepackage{calc}
\usepackage{color}
\usepackage{url}

\theoremstyle{plain}
\newtheorem{lem}{Lemma}
\newtheorem{thm}{Theorem}
\theoremstyle{definition}
\newtheorem*{defn}{Definition}
\theoremstyle{remark}

\makeatletter 

\newcommand{\seclbl}[1]{\label{sec:#1}}
\newcommand{\secref}[1]{\ref{sec:#1}}
\newcommand{\eqnlbl}[1]{\label{eqn:#1}}
\newcommand{\eqnref}[1]{\eqref{eqn:#1}}
\newcommand{\figlbl}[1]{\label{fig:#1}}
\newcommand{\figref}[1]{\ref{fig:#1}}

\newcommand{\beq}{\@ifstar\@@beq\@beq}
\newcommand{\@@beq}{}
\newcommand{\@beq}[1][]{\begin{equation}\ifx\\#1\\
\else\eqnlbl{#1}\fi}
\newcommand{\eeq}{\@ifstar\@@eeq\@eeq}
\newcommand{\@@eeq}{}
\newcommand{\@eeq}{\end{equation}}

\newcommand{\inputfig}{\@ifstar\@@inputfig\@inputfig}
\newcommand{\@@inputfig}[2][1]{\begin{center}\includegraphics[scale=#1]{#2}\end{center}}
\newcommand{\@inputfig}[3][1]{\begin{figure}[htbp]\centering\includegraphics[scale=#1]{#2}\caption{#3}\figlbl{#2}\end{figure}}

\newcommand{\setdef}{\@ifstar\@@setdef\@setdef}
\newcommand{\@@setdef}[1]{\left\lbrace #1 \right\rbrace}
\newcommand{\@setdef}[2]{\left\lbrace #1 \middle|\, #2 \right\rbrace}

\newcommand{\ci}{\@ifstar\@@ci\@ci}
\newcommand{\@@ci}{\mathrm i \,}
\newcommand{\@ci}{\mathrm i}

\newcommand{\li}{\@ifstar\@@li\@li}
\newcommand{\@@li}[1]{_{(#1)}}
\newcommand{\@li}[1]{_\mathrm{#1}}
\newcommand{\ui}{\@ifstar\@@ui\@ui}
\newcommand{\@@ui}[1]{^{(#1)}}
\newcommand{\@ui}[1]{^\mathrm{#1}}

\newcommand{\fn}{\@ifstar\@@fn\@fn}
\newcommand{\@@fn}[2]{\mathrm{#1}#2} 
\newcommand{\@fn}[2]{\mathrm{#1}\, #2}
\newcommand{\fnb}{\@ifstar\@@fnb\@fnb}
\newcommand{\@@fnb}[2]{\mathsf{#1}(#2)}
\newcommand{\@fnb}[2]{\mathrm{#1}(#2)}

\title{Mermin pentagrams arising from Veldkamp lines for three qubits}
\author{Péter Lévay and Zsolt Szabó}

\usepackage{hyperref}
\hypersetup{
	pdfauthor={\@author},
	pdftitle={\@title},
	pdfproducer={LaTeX with hyperref},
	pdffitwindow=false,
	pdfstartview={FitH},
	colorlinks=true,
	linkcolor={blue},
	citecolor={blue},
	filecolor={blue},
	urlcolor={blue}}

\begin{document}
\begin{center}
{\Large\bf\@title}
\end{center}
\vspace*{-.1cm}
\begin{center}
P\'eter L\'evay$^{1,2}$ and Zsolt Szab\'o$^{1}$
\end{center}
\vspace*{-.4cm}
\begin{center}
$^{1}$Department of Theoretical Physics, Institute of Physics,
Budapest University of \\
Technology and Economics 

$^{2}$MTA-BME Condensed Matter Research Group, H-1521 Budapest, Hungary

\vspace*{.0cm}
\vspace*{.2cm} (\today)
\end{center}
\makeatother


\vspace*{-.3cm} \noindent \hrulefill

\vspace*{.1cm} \noindent {\bf Abstract:}
We study the geometry of the space of Mermin pentagrams, objects that are used to rule out the existence of noncontextual hidden variable theories as alternatives to quantum theory. It is shown that this space of 12\,096 possible pentagrams is organized into 1008 families, with each family containing a "double-six" of pentagrams. The 1008 families are connected to a special set of Veldkamp lines in the Veldkamp space for three-qubits an object well-known to finite geometers but has only been introduced to physics recently. Due to the transitive action of the symplectic group on this set of Veldkamp lines it is enough to study only one "canonical" double-six configuration of pentagrams. We prove that the geometry of this double-six configuration is encapsulated in the weight diagram of the $20$ dimensional irreducible representation of the group $\fnb{SU}{6}$. As an interesting by-product of our approach we show that Mermin pentagrams and a class of Mermin squares labelled by three-qubit Pauli operators are inherently related. We conjecture that by studying the representation theoretic content of other Veldkamp lines of the Veldkamp space for $N$-qubits makes it possible to find new contextual configurations in a systematic manner.


\vspace*{.3cm} \noindent

\vspace*{-.2cm} \noindent \hrulefill


\section{Introduction}
In the 1990s there was a flurry of activity on revisiting the Kochen-Specker\cite{KS} and Bell theorems\cite{Bell}.
Peres\cite{Peres}, Mermin\cite{Mermin1990, Mermin1993} and Greenberger, Horne, Zeilinger\cite{Horne} have given elegant proofs of these theorems using systems of two, three and four qubits. The remarkable feature appearing in
these works was that they were able to rule out
certain classes of hidden variable theories without the use of probabilities.
Since the advent of quantum information theory these ideas have been again under an intense 
scrutiny\cite{Waegell,Waegell3qbit,Saniga1,Levay1,Planat,SLev,Levfin,BHQC}.
Apart from obtaining new proofs for these theorems\cite{Waegell,Waegell3qbit} some of these works also attempted to relate
the geometric configurations underlying these proofs to finite geometric structures well-known to
mathematicians\cite{Saniga1,Levay1,Planat,SLev}, or even to the structure of entropy formulas of some black hole solutions in string theory\cite{Levfin,Levfin2,BHQC}.
The basic idea of these considerations was to label the points of these configurations with the observables of some 
{\it finite} dimensional multipartite system
in a manner compatible with a set of rules encapsulating the commutation properties and some identities for the observables. 
The net result of this process was an incidence structure which encapsulated the inherent noncommutativity underlying our multiparite quantum system.

For $N$-qubit systems an approach of that kind was initiated in \cite{San1} with the incidence structure arising called
${\mathcal W}_{2N-1}(2)$ the symplectic polar space of rank $N$ order two\cite{Buek}. 
This space is equipped with a symplectic form with values in the two element field ${\mathbb Z}_2$, corresponding to the two possibilities for the observables are either commuting or not commuting.
The symplectic group $\fnb{Sp}{2N,{\mathbb Z}_2}\equiv \fnb{Sp}{2N,2}$ is the one leaving invariant the symplectic form (null-polarity) of ${\mathcal W}_{2N-1}(2)$.
The main application of this space is within the field of quantum information and it is related to quantum error correcting codes\cite{Nielsen}. 
The construction of such codes is naturally facilitated within the so called stabilizer formalism\cite{Nielsen,Gottesman,Sloane}. 
Here it is recognized that the basic properties of error correcting codes are related again to the fact that two operators in the Pauli group are either commuting or anticommuting.

Later it has also been realized that certain subconfigurations 
of ${\mathcal W}_{2N-1}(2)$
called geometric hyperplanes\cite{Shult} are also worth studying. For example in the case of ${\mathcal W}_3(2)$ one particular class of its geometric hyperplanes features the $10$ possible Mermin squares\cite{Mermin1990, Mermin1993} one can construct from two-qubit Pauli operators.
For mathematicians it is well-known that sometimes the set of geometric hyperplanes can be organized to Veldkamp lines which are in turn organized into a new incidence structure the so-called Veldkamp space\cite{Veld,Shult}.
In this spirit for the simplest nontrivial case of ${\mathcal W}_3(2)$ the structure of the Veldkamp space has been thoroughly investigated, 
the physical meaning of the geometric hyperplanes clarified, and pictorially illustrated\cite{San2}.
Since for an arbitrary number of qubits the diagrammatic approach of \cite{San2} is not feasible, a later study\cite{VranLev} has revealed the structure of the Veldkamp space of ${\mathcal W}_{2N-1}(2)$ in a purely algebraic fashion.

The $N=3$ (three-qubits) case is particularly interesting.
In this case the symplectic group $\fnb{Sp}{6,2}$ associated with ${\mathcal W}_5(2)$ is isomorphic\cite{Bour} to $W(\fn* E_7)/{\mathbb Z}_2$ 
with $W(\fn* E_7)$ the Weyl group of the exceptional group $\fn* E_7$.
This group is the automorphism group of a point-line incidence geometry ${\mathcal G}_3$ where the $63$ points are the points of $W_5(2)$ and the $315$ lines are the totally isotropic ones of ${\mathcal W}_5(2)$ with respect to the symplectic form\cite{VranLev}.
A nice way of understanding this is to note that there is a bijection\cite{Geemen} between the $63$ points of ${\mathcal W}_5(2)$ and the 63 pairs of roots of $\fn* E_7$.
One particular type of geometric hyperplanes of ${\mathcal G}_3$ is featuring $27$ points and $45$ lines and having the incidence geometry of a generalized quadrangle\cite{Payne} $\fnb{GQ}{2,4}$ with the automorphism group $W(\fn* E_6)$. In \cite{Levfin} it has been shown
how these structures encode information on the structures of the $\fn* E_{7(7)}$ and $\fn* E_{6(6)}$ symmetric black hole entropy formulas arising in certain models of string theoretic compactifications. 
It has been also observed\cite{Levfin2,BHQC} that certain truncations of these models result in truncations of the corresponding entropy formulas, which in turn correspond to truncations to further geometric hyperplanes.
For example the $27$ points of $\fnb{GQ}{2,4}$ can be partitioned into three sets of Mermin {\it squares} with $9$ points each.
This partitioning corresponds to the reduction of the $27$ dimensional irreducible representation of $\fn* E_{6(6)}$ to a substructure arising from three copies of $3$ dimensional irreps of three $\fnb{SL}{3,{\mathbb R}}$s.
The configuration related to this truncation has an interesting physical interpretation in terms of wrapped membrane configurations and is known in the literature as the bipartite entanglement of three qutrits\cite{Duff,BHQC}.
These results hint at a deep connection between the structure of geometric hyperplanes (\emph{Veldkamp points}), contextual configurations\cite{HS} and the structure of irreducible representations of certain types of Lie-groups. 

In this paper we reveal a connection between the set of \emph{Mermin pentagrams} and a particular class of \emph{Veldkamp lines} for three qubits. The number of Mermin pentagrams is 12\,096 (see \cite{Saniga1}), and these organize into 1008 families of 12 pentagrams\cite{Zsolt}. We call these families \emph{double sixes of pentagrams}.
We would like to understand the geometry of these spaces.
We show that each family  is inherently connected to the structure of the weight diagram for the $20$ dimensional irrep of $\fnb{SU}{6}$.
It turns out that there are $1008$ such families that can be mapped bijectively to a subclass of Veldkamp lines of the Veldkamp space for three-qubits.
Since there is a transitive action of the symplectic group $\fnb{Sp}{6,2}$ on this class of Veldkamp lines\cite{VranLev} it is enough to study merely one particular family, which we call canonical.
As a byproduct of our approach we also establish an interesting connection between a subset of the space of Mermin {\it squares} formed from three-qubit operators and the families of double-sixes of Mermin {\it pentagrams}.

The organization of this paper is as follows.
In Section \secref{mathbg} we collect the relevant background information on the $N$-qubit Pauli group related to the structure of the symplectic polar space ${\mathcal W}_{2N-1}(2)$. Here our basic notions like geometric hyperplanes, Lagrange subspaces, Veldkamp lines and Veldkamp spaces for $N$-qubits are defined.
In Section \secref{double_six} the space of Mermin pentagrams for three-qubit systems is introduced.
Here we also introduce the notion of a double-six configuration of 12 pentagrams.
In Section \secref{doily_spreads} we demonstrate the connection between the structure of this double-six configuration of Mermin pentagrams and the structure of the spreads of the generalized quandrangle $\fnb{GQ}{2,2}$ also called the \emph{doily}\cite{Payne,Polster}. Armed with this observation, in Section \secref{Veldkamp_lines} we easily prove that a particular set of $1008$ Veldkamp lines with their core sets being $1008$ doilies encode all relevant information concerning the space of Mermin pentagrams.
In fact we have to study merely one particular Veldkamp line, with a particular doily and a particular double-six.
The reason for this is that the symplectic group acts transitively\cite{VranLev} on our particular $1008$ element set of Veldkamp lines. Hence by uncovering the structure of merely one canonical arrangement for the aforementioned objects we have uncovered the structure of all such ones. 
In Section \secref{representation} we prove that the geometry of the canonical double-six configuration (hence by transitivity all of such ones) is encapsulated entirely in the weight diagram of the $20$ dimensional irreducible representation of the group $\fnb{SU}{6}$.
Section \secref{conclusions} is left for the conclusions and some speculations.
Here we notice that a certain $10\,080$ element subset of the space of Mermin squares is related to the $12\,096$ element set of Mermin pentagrams in a simple manner. Here we also conjecture that the structure of Veldkamp lines for $N$-qubits ($N=2,3,4$) is inherently connected to special irreducible representations of ADE-type Lie algebras. This observation might be a starting point for obtaining new magic configurations in a systematic manner. 


\section{Mathematical background}
\seclbl{mathbg}

We start by recapitulating some basic notions and facts about the building blocks of the set of Mermin pentagrams. The \emph{observables} of a quantum system described by a finite dimensional Hilbert space $\mathcal H$ are represented by self-adjoint operators on $\mathcal H$, and if a basis is fixed in $\mathcal H$, each observable is equivalent to a Hermitian matrix. In the case of a two-state quantum system, or \emph{qubit}, a basis for the vector space of the $2\times 2$ Hermitian matrixes is $\setdef*{I,X,Z,Y}$, where $I$ is the $2\times 2$ identity matrix, and
\beq[paulimat]
X=\begin{pmatrix}
0 & 1 \\
1 & 0
\end{pmatrix},
\qquad
Z=\begin{pmatrix}
1 & 0 \\
0 & -1
\end{pmatrix},
\qquad
Y = \ci XZ =\begin{pmatrix}
0 & -\ci \\
\ci & 0
\end{pmatrix} \eeq
are the three \emph{Pauli spin matrices}. The matrices of a special set of observables of a system of $N$ qubits are the $N$-fold Kronecker products of $I$, $X$, $Z$ and $Y$. We will refer to these $2^N\times 2^N$ matrices as \emph{$N$-qubit Pauli operators}.

The $N$-qubit Pauli operators generate the \emph{$N$-qubit Pauli group}
\beq P_N = \setdef{s A_1 A_2 ... A_N}{s\in\mu_4, \, A_i = I,X,Z,Y} \qquad \leq GL(2^{N},\mathbb C), \eeq
where $A_1 A_2 ... A_N$ is a shorthand notation for $A_1\otimes A_2 \otimes\cdots\otimes A_N$, and $\mu_4$ is the group of the fourth roots of unity:
\beq \mu_4 = \setdef*{\pm 1, \pm\ci} \subseteq \mathbb C \setminus \setdef* {0}. \eeq

The commutator subgroup of $P_N$, i.e.\ the derived group $P_N' = [P_N, P_N]$ coincides with the center of $P_N$,
\beq[zpn] Z(P_N) = \setdef{s I_N}{s\in\mu_4}, \eeq
where $I_N$ is the $2^N\times 2^N$ identity matrix. It follows that the central quotient $P_N / Z(P_N)$ is an abelian group. The elements of $P_N / Z(P_N)$ are equivalence classes of the form
\beq \setdef{s A_1 A_2 ... A_N}{s\in\mu_4} = \setdef*{\pm A_1 A_2 ... A_N, \pm\ci A_1 A_2 ... A_N}, \eeq
and from each class we choose the $N$-qubit Pauli operator $A_1 A_2 ... A_N$ as the representative.

Now we specialize to $P_3$, but we note that most of our remarks can be easily reformulated for the general case. We are mainly interested in two aspects of $P_3$. First, $P_3$ has the structure of a symplectic vector space over $\mathbb Z_2$. Second, $P_3$ induces a symplectic polar space $W_5(2)$, which eventually leads us to what we call the \emph{three-qubit Veldkamp space} whose special class of lines is the main concern of this paper.

Let us start with the symplectic vector space structure of $P_3$. By \eqnref{paulimat}, an arbitrary group element $A = s A_1 A_2 A_3 \in P_3$ can be written as
\beq[tprod]
A = \left( \ci^{\sum_i a_i b_i} \right) \, s X^{a_1}Z^{b_1}\otimes X^{a_2}Z^{b_2}\otimes X^{a_3}Z^{b_3},
\eeq
where the exponents of the Pauli matrices take their value from $\setdef*{0,1}\subseteq\mathbb N$. \eqnref{tprod} shows that $A$ can be uniquely identified by the 7-tuple
\beq[7tuple] (s, a_1, a_2, a_3, b_1, b_2, b_3). \eeq
By discarding the first element of \eqnref{7tuple}, we get a vector
\beq x = (a_1, a_2, a_3, b_1, b_2, b_3) \in \mathbb Z_2^6, \eeq
that identifies the 3-qubit Pauli operator $A_1 A_2 A_3$, and consequently, the respective equivalence class of $P_3 / Z(P_3)$. Because of $ZX = -XZ$ and $X^2 = Z^2 = I$, in the case of the matrix product $AA'$ of $A,A' \in P_3$, \eqnref{7tuple} becomes
\beq[prod7tuple] \left( ss' \, (-1)^{\sum_i a_i'b_i},a_1+a_1',\dots, b_3+b_3' \right), \eeq
where ``+'' means addition modulo 2, which is the addition of $\mathbb Z_2$. Moreover, \eqnref{prod7tuple} implies that the group multiplication of $P_3 / Z(P_3)$ induces the vector addition in $\mathbb Z_2^6$.

\eqnref{prod7tuple} also implies that $A$ and $A'$ commute iff
\beq[komm] \sum_{i=1}^{3}(a_i b_i' + a_i' b_i) = 0. \eeq
The left hand side of \eqnref{komm} is a non-degenerate, skew-symmetric bilinear form on $\mathbb Z_2^6$. In general, such bilinear forms are called \emph{symplectic forms}, and vector spaces equipped with a symplectic form are called \emph{symplectic vector spaces}. Thus $\mathbb Z_2^6$ with the symplectic form
\beq[simplform]
\langle\cdot,\cdot\rangle: \mathbb Z_2^6 \times \mathbb Z_2^6 \to {\mathbb Z}_2, \qquad (x,x') \mapsto \langle x,x^{\prime}\rangle\equiv\sum_{i=1}^3 (a_i b_i' + b_i a_i'). \eeq
is a symplectic vector space over $\mathbb Z_2$, which in the following will be denoted by $V_3$. By a slight abuse of notation, we state the meaning of $\langle\cdot,\cdot\rangle$ as follows: given two equivalence classes of $P_3 / Z(P_3)$ identified by $x,x'\in V_3$, an arbitrary element $A\in x$ commutes with an arbitrary element $A'\in x'$ iff $\langle x, x' \rangle = 0$. Often it is more convenient to write the vectors of $V_3$ as three-qubit Pauli operators instead of 6-tuples of 0's and 1's. For example, by \eqnref{tprod}, $XYZ$ is equivalent to $(1,1,0,0,1,1)$.

$V_3$ has special bases, also called \emph{symplectic}, that are analogous to orthonormal bases of Euclidean spaces. Without loss of generality, we denote a symplectic basis of $V_3$ by
\beq[sympbasis] \{ e_1, e_2, e_3, f_1, f_2, f_3 \}, \eeq
and we assume that
\beq \langle e_i, e_j \rangle = \langle f_i, f_j \rangle = 0, \qquad \langle e_i, f_j \rangle = \delta_{ij}, \qquad i,j=1,2,3. \eeq
The most important example of a symplectic basis is the \emph{canonical basis} of $V_3$:
\beq[can_basis]
\begin{aligned}
e_1 = XII &= (1,0,0,0,0,0), \qquad & f_1 = ZII &= (0,0,0,1,0,0), \\
e_2 = IXI &= (0,1,0,0,0,0), \qquad & f_2 = IZI &= (0,0,0,0,1,0), \\
e_3 = IIX &= (0,0,1,0,0,0), \qquad & f_3 = IIZ &= (0,0,0,0,0,1).
\end{aligned}\eeq
By letting $e_\mu = f_{\mu-3}$ for $\mu = 4,5,6$, the matrix of the symplectic form with respect to the canonical basis can be written as
\beq[simplmatr]
J_{\mu\nu}\equiv \langle e_{\mu},e_{\nu}\rangle = \begin{pmatrix}
0 & I_3 \\
I_3 & 0
\end{pmatrix}, \qquad \mu,\nu=1,2,\dots 6. \eeq
where $0$ is the $3\times 3$ zero matrix, and $I_3$ is the $3\times 3$ identity matrix.

The invariance group of the symplectic form $\langle\cdot,\cdot\rangle$ is the group with matrix multiplication as group operation of the $6\times 6$ matrices $S$ that preserve $\langle\cdot,\cdot\rangle$, and consequently satisfy
\beq[sympgdef]
S\ui T J S = J.
\eeq
This group is called the \emph{symplectic group of order 6 over $\mathbb Z_2$}, and it is denoted by
\beq \fnb{Sp}{6,\mathbb Z_2} \equiv \fnb{Sp}{6,2}. \eeq
For each vector $p\in V_3$ there exists a \emph{transvection} $T_p\in \fnb{Sp}{6,2}$ that acts on $V_3$ as follows:
\beq[transv]
T_p x = x+\langle x,p\rangle p, \qquad x\in V_3.
\eeq
It is known that the set of transvections generates $\fnb{Sp}{6,2}$ \cite{SpGen} and that there is a surjective homomorphism \cite{Bour} from $W(\fn* E_7)$ i.e.\ the Weyl group of the exceptional group $\fn* E_7$ to $\fnb{Sp}{6,2}$ with kernel ${\mathbb Z}_2$.

Next, we turn to the symplectic polar space realized by $P_3$, and to its geometric hyperplanes, but first we need a few general definitions.


\begin{defn}
An \emph{incidence structure} (or \emph{point-line incidence geometry}) is a triple $({\mathcal P},{\mathcal L},{\mathcal I})$, where ${\mathcal P}$ and ${\mathcal L}$ are disjoint sets and ${\mathcal I}\subseteq {\mathcal P}\times {\mathcal L}$ is a relation, called the \emph{incidence relation}. The elements of ${\mathcal P}$ and ${\mathcal L}$ are called \emph{points} and \emph{lines} respectively. We say that a point $p\in {\mathcal P}$ is \emph{incident} with a line $l\in {\mathcal L}$ if $(p,l)\in {\mathcal I}$, and two points $p,p'\in\mathcal P$ are \emph{collinear}, if both are incident with some line $l\in \mathcal L$.
\end{defn}
In the following we limit ourselves to incidence structures whose lines are realized by subsets of the point set $\mathcal P$, and whose incidence relation is $\in$. In other words, $\mathcal L\subseteq 2^{\mathcal P}$, and $p\in\mathcal P$ is incident with $l\in\mathcal L$ iff $p\in l$. Such incidence structures $({\mathcal P},{\mathcal L},\in)$ are called \emph{simple}.

Next, we define special sets of points of an incidence structure, called geometric hyperplanes, that will be important later.

\begin{defn}
A subset ${\mathcal H}\subseteq {\mathcal P}$ is a \emph{geometric hyperplane} of the incidence structure $({\mathcal P},{\mathcal L},\in)$ if the following two conditions hold \cite{Shult}:
\begin{enumerate}[(H1)]
\item $l\subseteq\mathcal H$ or $|\mathcal H\cap l| = 1$ for all $l\in \mathcal L$;
\item $\mathcal H \neq \mathcal P$.
\end{enumerate}
\end{defn}

Our aim is to examine the incidence structure that emerges from the \emph{$N$-qubit Pauli group} $P_N$, and to study its geometric hyperplanes.

The projective space $PG(2N-1,2)$ consists of the nonzero subspaces of the $2N$ dimensional vector space $V_N$ over $\mathbb{Z}_2$. The points of the projective space are one dimensional subspaces of the vector space, and more generally, $k$ dimensional subspaces of the vector space are $k-1$ dimensional subspaces of the corresponding projective space. A subspace of $(V_N,\langle\cdot,\cdot\rangle)$  (and also the subspace in the corresponding projective space) is called \emph{isotropic} if there is a vector in it which is orthogonal (with respect to the symplectic form) to the whole subspace, and \emph{totally isotropic} if the subspace is orthogonal to itself. The space of totally isotropic subspaces of $(PG(2N-1,2),\langle\cdot,\cdot\rangle)$ is called the {\it symplectic polar space} of rank $2N-1$ order two denoted by ${\mathcal W}(2N-1,2)$. The maximal totally isotropic subspaces are called \emph{Lagrangian subspaces}.

Let us illustrate these abstract concepts in terms of the physically relevant structures of three-qubit observables. The $63$ nonzero vectors in $V_3$ are comprising the points of $PG(5,2)$. Modulo an element of $\mu_4$ they correspond to the nontrivial observables (i.e.\ the ones excluding the identity $III$). Let $W$ be some subspace of $PG(5,2)$ corresponding to a subset of nontrivial observables.
Then
$W^{\perp}$ the orthogonal complement with respect to the symplectic form corresponds to the set of the observables {\it commuting} with the ones in $W$. A line $L\in PG(5,2)$ is an object of the form $L=\lambda v+\mu u$ where $u,v\in PG(5,2)$ and $\lambda,\mu\in {\mathbb Z}_2$ containing the three points: $u,v,u+v$. Hence a line corresponds to a triple of operators such that any two of them multiplto the third (up to an element of $\mu_4$). An example of a line represented by three-qubit operators is $\{ IIX, IIY, IIZ \}$. Now, the representative operators of a totally isotropic line, e.g. $\{ IIX, IXI, IXX \}$, are also \emph{pairwise commuting}. The number of totally isotropic lines is $315 = (64-1)(32-2)/6$.

A totally isotropic plane in $PG(5,2)$ is of the form $P=\lambda v+\mu u+\nu w$ where $u,v,w$ are linearly independent and pairwise commuting. Clearly, $P$ contains seven points and seven totally isotropic lines. Since each line features three points and each point is incident with three lines, the incidence structure of the Fano plane, or $PG(2,2)$ emerges\cite{Polster}. As an example of a seven-tuple comprising the points of a Fano plane one can take the pairwise commuting set $\{IIX,IXI,IXX,XII,XIX,XXI,XXX\}$. The totally isotropic planes in $PG(5,2)$ are maximal hence they are Lagrangian subspaces.
It was shown\cite{Geemen} that the number of Lagrangian subspaces (which are in turn Fano planes) is 135. The geometry of the set of Lagrangian subspaces for three-qubits has been explored in\cite{Levay1}. In \cite{Levay1} 
in terms of the generators of a seven dimensional Clifford algebra
an explicit list of all 135 Lagrangian subspaces was also presented.
In summary: we see that ${\mathcal W}(5,2)$ is comprising the $63$ points the $315$ totally isotropic lines and $135$ totally isotropic planes of $PG(5,2)$.

For an element $x\in V_3$ let us define the quadratic form
\beq[kankvadrat] Q_0\equiv \sum_{i=1}^3 a_ib_i. \eeq
It is easy to check that for vectors representing symmetric operators $Q_0(x)=0$ (the ones containing an {\it even} number of $Y$s) and for antisymmetric ones $Q_0(x)=1$ (the ones containing an {\it odd} number of $Y$s). Moreover we have the relation 
\beq[kvdratosszef] \langle x,y\rangle =Q_0(x+y)+Q_0(x)+Q_0(y). \eeq
The \eqnref{kankvadrat} quadratic form will be regarded as a one labelled by the trivial element of $V_3$ with representative $III$. There are however, $63$ other quadratic forms $Q_p$ compatible the symplectic form $\langle\cdot\cdot\rangle$ labelled by the nontrivial elements $p$ of $V_3$ also satisfying
\beq \langle x,y\rangle =Q_p(x+y)+Q_p(x)+Q_p(y). \eeq
They are defined as
\beq[ujkvadforms] Q_p(x)=Q_0(x)+\langle x,p\rangle^2 \eeq
since we are over the two element field the square can be omitted. 

For more information on these quadratic forms we orient the reader to \cite{Geemen,VranLev}. Here we merely elaborate on the important fact that there are two classes of such quadratic forms. They are the ones that are labelled by symmetric observables ($Q_0(p)=0$), and antisymmetric ones ($Q_0(p)=1$). The locus of points in $PG(5,2)$ satisfying $Q_p(x)=0$ for $Q_0(p)=0$ is called a {\it hyperbolic} quadric and the locus $Q_p(x)=0$ for which $Q_0(p)=1$ is called an {\it elliptic} one. The former one will be denoted by $Q^+(5,2)$ and the latter by $Q^-(5,2)$. Looking at \eqnref{ujkvadforms} one can see that in terms of three-qubit observables (modulo elements of $\mu_4$) one can characterize the quadrics $Q(5,2)$ as follows. The three-qubit observables $x\in Q(5,2)$ characterized by $Q_p(x)=0$ are the ones that are either symmetric and commuting with $p$ or antisymmetric and anticommuting with $p$. It can be shown\cite{VranLev,Geemen} that we have $36$ quadrics of type $Q^+(5,2)$ and $28$ ones of type $Q^-(5,2)$, with the former containing $35$ and the latter containing $27$ points of $PG(5,2)$. A hyperbolic quadric of $Q^+(5,2)$ type in $PG(5,2)$ is called the \emph{Klein-quadric}. On the other hand an elliptic quadric of $Q^-(5,2)$ type can be shown to display the structure of a generalized quadrangle\cite{Payne} $\fnb{GQ}{2,4}$ an object we already mentioned in the introduction. A pictorial representation of this incidence structure having $27$ points and $45$ lines labelled by three-qubit observables can be found in \cite{Levfin2}.

Now we are ready to present the point-line incidence structure of $N$-qubit observables.

\begin{defn}
Let $N\in\mathbb{N}+1$ be a positive integer, and $V_{N}$ be the symplectic $\mathbb{Z}_{2}$-linear space. The incidence structure $\mathcal{G}_{N}$ of the $N$-qubit Pauli group is $({\mathcal P},{\mathcal L},\in)$ where ${\mathcal P}=V_{N}\setminus\{0\}$,
\beq {\mathcal L}=\{\{a,b,a+b\}|a,b\in {\mathcal P},a\neq b,\langle a,b\rangle=0\} \eeq
and $\in$ is the set theoretic membership relation.
\end{defn}

Clearly the points and lines of $\mathcal{G}_{N}$ are the ones of the symplectic polar space ${\mathcal W}(2N-1,2)$. Of course our main concern here is the $N=3$ case. In this case $\mathcal{G}_{3}$ has $63$ points and $315$ lines.

Our next task is to recall the properties of the geometric hyperplanes of $\mathcal{G}_{N}$. The following lemma was proved in \cite{VranLev}.

\begin{lem}\label{lem:hypplanes}
Let $N\in\mathbb{N}+1$ be a positive integer, $\mathcal{G}_{N}=({\mathcal P},{\mathcal L},\in)$ and $p\in V_{N}$ be any vector. Then the following sets satisfy (H1):
\begin{align}
C_{p} &= \{x\in {\mathcal P}|\langle p,x\rangle=0\}, \\
H_{p} &= \{x\in {\mathcal P}|Q_{p}(x)=0\}.
\end{align}
\end{lem}

This lemma shows that apart from $C_0$ all of the sets above are geometric hyperplanes of the geometry $\mathcal{G}_{N}$. The set $C_p$ is called the {\it perp set} of $p\in V_N$. Modulo an element of $\mu_4$ $C_p$ represents the set of observables commuting with a fixed one $p$. Back to the implications of our lemma one can show that in fact more is true, {\it all geometric hyperplanes} arise in this form\cite{VranLev}:
\begin{thm}
Let $N\in\mathbb{N}+1$, $\mathcal{G}_{n}=({\mathcal P},{\mathcal L},\in)$, and ${\mathcal H}\in {\mathcal P}$ a subset
satisfying (H1). Then either ${\mathcal H}=C_{p}$ or ${\mathcal H}=H_{p}$ for some $p\in V_{N}$.
\end{thm}

One can prove that for $N\geq 2$ no geometric hyperplane is contained in another one, more precisely\cite{VranLev}
\begin{thm}\label{thm:Vpoints}
Let $N\in\mathbb{N}+2$, $\mathcal{G}_{n}=({\mathcal P},{\mathcal L},\in)$ and suppose that $A,B\subset {\mathcal P}$ are two geometric hyperplanes. Then $A\subseteq B$ implies $A=B$.
\end{thm}
Another property of two different geometric hyperplanes is that the complement of their symmetric difference gives rise to a third geometric hyperplane i.e.\

\begin{lem}
For $A\neq B$ geometric hyperplanes in $\mathcal{G}_{N}=({\mathcal P},{\mathcal L},\in)$
with $N\ge 1$ the set
\beq A\boxplus B:=\overline{A\triangle B}=(A\cap B)\cup(\overline{A}\cap\overline{B}) \eeq
is also a geometric hyperplane.
\end{lem}
One can also check that by using the notation $C\equiv A\boxplus B$
\beq
A\cap C = A\cap B,\qquad B\cap C= A\cap B, \qquad A\boxplus C = B. \eeq
A corollary of this is that any two of the triple $(A,B,A\boxplus B)$ of hyperplanes determines the third.

Sometimes it is also possible to associate to a particular incidence geometry another one called its Veldkamp space whose points are geometric hyperplanes of the original geometry\cite{Shult}:
\begin{defn}
Let $\Gamma=({\mathcal P},{\mathcal L},{\mathcal I})$ be a point-line geometry. We say that $\Gamma$ has \emph{Veldkamp points} and \emph{Veldkamp lines} if it satisfies the conditions
\begin{enumerate}[(V1)]
\item For any hyperplane $A$ it is not properly contained in any other hyperplane $B$.
\item For any three distinct hyperplanes $A$, $B$ and $C$, $A\cap B\subseteq C$ implies $A\cap B=A\cap C$.
\end{enumerate}
If $\Gamma$ has Veldkamp points and Veldkamp lines, then we can form the
Veldkamp space $V(\Gamma)=({\mathcal P}_{V},{\mathcal L}_{V},\supseteq)$ of $\Gamma$, where ${\mathcal P}_{V}$ is the set of geometric hyperplanes of $\Gamma$, and ${\mathcal L}_{V}$ is the set of intersections of pairs of distinct hyperplanes.
\end{defn}

Clearly, by Theorem \ref{thm:Vpoints}, $\mathcal{G}_{N}$ contains Veldkamp points for $N\geq 2$ hence in this case V1 is satisfied. In order to see that V2 holds as well we note\cite{VranLev}
\begin{lem}\label{lem:Vlines}
Let $N\in\mathbb{N}+1$, $a,b\in V_{N}$ and $\mathcal{G}_{N}=({\mathcal P},{\mathcal L},\in)$. Then the
following formulas hold:
\begin{align}
\begin{split}
C_{a}\boxplus C_{b} &= C_{a+b}, \\
H_{a}\boxplus H_{b} &= C_{a+b}, \\
C_{a}\boxplus H_{b} &= H_{a+b}.
\end{split}
\end{align}
\end{lem}
From this it follows that for any three geometric hyperplanes $A,B,C$ we have $A\cap B=A\cap C=B\cap C$.
One can however show more\cite{VranLev}, namely that there is no other possibility i.e.\  $A\cap B\subseteq C$ implies $C\in\{A,B,A\boxplus B\}$.
\begin{thm}
Let $N\in\mathbb{N}+3$, and suppose that $A,B,C$ are distinct geometric
hyperplanes of $\mathcal{G}_{n}=({\mathcal P},{\mathcal L},\in)$ such that $A\cap B\subseteq C$.
Then $A\cap B=A\cap C$.
\end{thm}
Notice that the statement is not true for $N=2$.

From these results it follows that there are two types of Veldkamp lines incident with three $C$-hyperplanes and three types of lines incident with one $C$-hyperplane and two $H$-hyperplanes. The former two types arise as $a$ and $b$ from $C_a$ and $C_b$ satisfy either $\langle a,b\rangle=0$ or $\langle a,b\rangle=1$. For the three types featuring two $H$-type hyperplanes are
\begin{eqnarray}
\setdef{ \{H_{a},H_{b}\} }{ a,b\in V_{n}, a\neq b, Q_0(a) = Q_0(b) = 0 }, \\
\setdef{ \{H_{a},H_{b}\} }{ a,b\in V_{n}, a\neq b, Q_0(a) = Q_0(b) = 1 }, \\
\setdef{ \{H_{a},H_{b}\} }{ a,b\in V_{n}, Q_{0}(a) \neq Q_{0}(b) }.
\end{eqnarray}
The third type will be of great importance for us. For $N=3$ this case gives rise to a Veldkamp line featuring a hyperbolic quadric, an elliptic quadric and a perp set. A canonical representation for this line can be chosen as $\{H_{III},H_{YYY},C_{YYY}\}$. An important result of \cite{VranLev} is the statement that $\fnb{Sp}{2N,2}$ acts transitively on each of the previous three sets of dublets of geometric hyperplanes of $H$-type. For $N=3$ we have 36 possibilities for choosing the hyperbolic and $28$ ones for choosing the elliptic one. Hence in the class with canonical representative $\{H_{III},H_{YYY},C_{YYY}\}$ we have altogether $36\cdot 28=1008$ Veldkamp lines. Then, by the transitivity of $\fnb{Sp}{6,2}$, we can reach any of the 1008 Veldkamp lines from the canonical one via a set of suitable symplectic transvections of the form \eqnref{transv}. For the construction of the explicit form of such transvections see \cite{VranLev}. A detailed description of these Veldkamp lines which comprise the Veldkamp space for the incidence structure ${\mathcal G}_N$ can be found in Table 2.\ of \cite{VranLev}. Here, apart from the list of Veldkamp lines, the Reader will find the cardinalities of the core sets (intersection sizes of the three hyperplanes featuring the Veldkamp lines) and the number of copies of a particular type of Veldkamp lines. Since in the following we only need the $N=3$ case and the aforementioned class of cardinality $1008$ , we refrain from presenting more details on these interesting issues.


\section{Mermin pentagrams and their double-sixes}
\seclbl{double_six}

\emph{Mermin's pentagram} is a set of ten observables, more specifically, nontrivial three-qubit Pauli-operators, that is used to prove the Bell--KS theorem in eight dimensions, i.e.\ to rule out non-contextual hidden-variables theories\cite{Mermin1993}. These observables are further arranged into five sets, or, as we like to call them, \emph{lines} (see `incidence structure' introduced in Section \secref{mathbg}).
\begin{enumerate}[(P1)]
\item Each line contains, or is incident with four pairwise commuting operators that multiply to $\pm III$, hence one can speak of \emph{positive} and \emph{negative lines}.
\end{enumerate}
Mermin's pentagram, published in \cite{Mermin1993}, is reproduced in Figure \figref{figure1}. The configuration has four positive lines and a negative one; in the figure the latter is highlighted in boldface.

\inputfig{figure1}{Mermin's pentagram}

The other properties relevant to the proof of the Bell--KS theorem are as follows\cite{Waegell3qbit}:
\begin{enumerate}[(P1)]
\stepcounter{enumi}
\item the number of lines each observable is incident with is even (two);
\item the number of negative lines is odd.
\end{enumerate}
By definition, a non-contextual hidden-variables theory is supposed to assign a value $v(A)$ to each possible observable $A$, such that $v(A)$ is an eigenvalue of $A$ and
\beq v(AB\cdots) = v(A) v(B) \cdots \eeq
holds for any pairwise commuting observables $A$, $B$, ... . Property (P1) means that for any line $\{ A_i, A_j, A_k, A_l \}$ of the pentagram we have
\beq A_i A_j A_k A_l = \pm III. \eeq
Since the only eigenvalue of the observable $\pm III$ is $\pm 1$, $v(\pm III) = \pm 1$. Therefore, the values $v(A_i)$,  $v(A_j)$, $v(A_k)$, $v(A_l)$ assigned to the observables of any line $\{ A_i, A_j, A_k, A_l \}$ of the pentagram must satisfy
\beq[mpentline] \pm 1 = v(\pm III) = v(A_i A_j A_k A_l) = v(A_i) v(A_j) v(A_k) v(A_l). \eeq
In the case of a positive line and a negative line, the LHS is $+1$ and $-1$, respectively. Now, consider the product of \eqnref{mpentline} for all five lines of the pentagram. The LHS multiplies to $-1$ due to (P3). The product of the RHS is of the form
\beq v(A_1)^{a_1} v(A_2)^{a_1} \cdots v(A_{10})^{a_{10}} = +1, \eeq
since $v(A_i) = \pm 1$, and since the exponents $a_i$ are even due to (P2). Thus we reach a contradiction.
For further details we refer the reader to \cite{Peresbook}, Chapter 7.


We call a \emph{Mermin pentagram}, or just simply a \emph{pentagram} any set of ten three-qubit Pauli operators that can be arranged as the observables shown in Figure \figref{figure1} and have properties (P1)-(P3) described above. Property (P2) is inherent to the pentagram-like arrangement. We know that (P3) follows from the rest\cite{HS}, and that there are $7884$, $4104$ and $108$ Mermin pentagrams with one, three and five negative lines respectively\cite{Saniga1}.

In Section \secref{mathbg} we saw that the three-qubit Pauli operators can be identified with the elements of the symplectic vector space $V_3$ over $\mathbb Z_2$ such that the multiplication of Pauli operators modulo $Z(P_3)$ (see \eqnref{zpn}) corresponds to the vector addition in $V_3$ (see \eqnref{prod7tuple}), and the commutation of two operators corresponds to orthogonality with respect to the symplectic form $\langle \cdot, \cdot \rangle$ of \eqnref{simplform}. In this Section and the next we will treat Pauli operators as vectors of $V_3$. Altough by doing this we cut ourselves off from discussing property P3 however, for Mermin pentagrams this property is automatically satisfied\cite{HS} for any consistent three-qubit labelling. 
Actually more is true: for a pentagram regarded as a subgeometry of a multi-qubit symplectic polar space any consistent labelling guarantees that there are an odd number of negative lines (contexts)\cite{HS}. 

An important observation\footnote{This follows from properties (P1) and (P2) of the pentagrams. The argument can be found in \cite{Zsolt}.} is that the six Pauli operators \emph{not} incident with an arbitrary line $l$ of a Mermin pentagram form a symplectic basis of $V_3$ (see \eqnref{sympbasis}). Moreover, since $l$ shares exactly one operator with each line $l'\neq l$, due to (P1) the shared operator is determined by the other three incident with $l'$. Thus pentagrams can be constructed from symplectic bases of $V_3$. However, this construction is not without ambiguity: each symplectic basis determines \emph{two} pentagrams. For instance, the pentagrams constructed from the canonical basis \eqnref{can_basis} are shown in Figure \figref{figure2}. Notice that the arrangements of the basis vectors differ only by an exchange of $XII$ and $ZII$. We say that these two pentagrams are each others \emph{conjugate}.

\inputfig{figure2}{Mermin pentagrams constructed from the canonical symplectic basis of $V_3$}

In general, every Mermin pentagram $P$ has five conjugates, and each conjugate is associated with the line of $P$ incident with the four observables not shared with the respective conjugate. Thus the space of Mermin pentagrams can be regarded as a regular graph of degree five, where the pentagrams play the r\^ole of vertices and two pentagrams are adjacent iff they are conjugates. A computer program prepared to explore this graph revealed 1008 connected components, all isomorphic to the intersection graph of Schl\"afli's double six configuration. The latter graph is shown in Figure \figref{figure3} (cf.\ \cite{Polster}, Section 4.5), and, to simplify the discussion, we will refer to \emph{it} as \emph{double six}.

\inputfig{figure3}{Intersection graph of Schl\"afli's double six, or `double six' in our terminology}

The double six is a bipartite graph with 12 vertices and 30 edges, and each vertex $v$ has an \emph{antipodal} vertex $v'$ such that $v$ and $v'$ are not adjacent and belong to different partitions. The double six of pentagrams that contains the two pentagrams in Figure \figref{figure2}, obtained from the canonical basis, can be seen in Figure \figref{figure4}. Instead of conjugates, Figure \figref{figure4} emphasizes the two partitions, the common sets of four observables forming a line (solid curves) and the antipodal pentagrams (the ones connected by dashed lines). In Figure \figref{figure4} we can see a total of 20 observables, and these are already contained by any pair of antipodal pentagrams.

\inputfig[0.96]{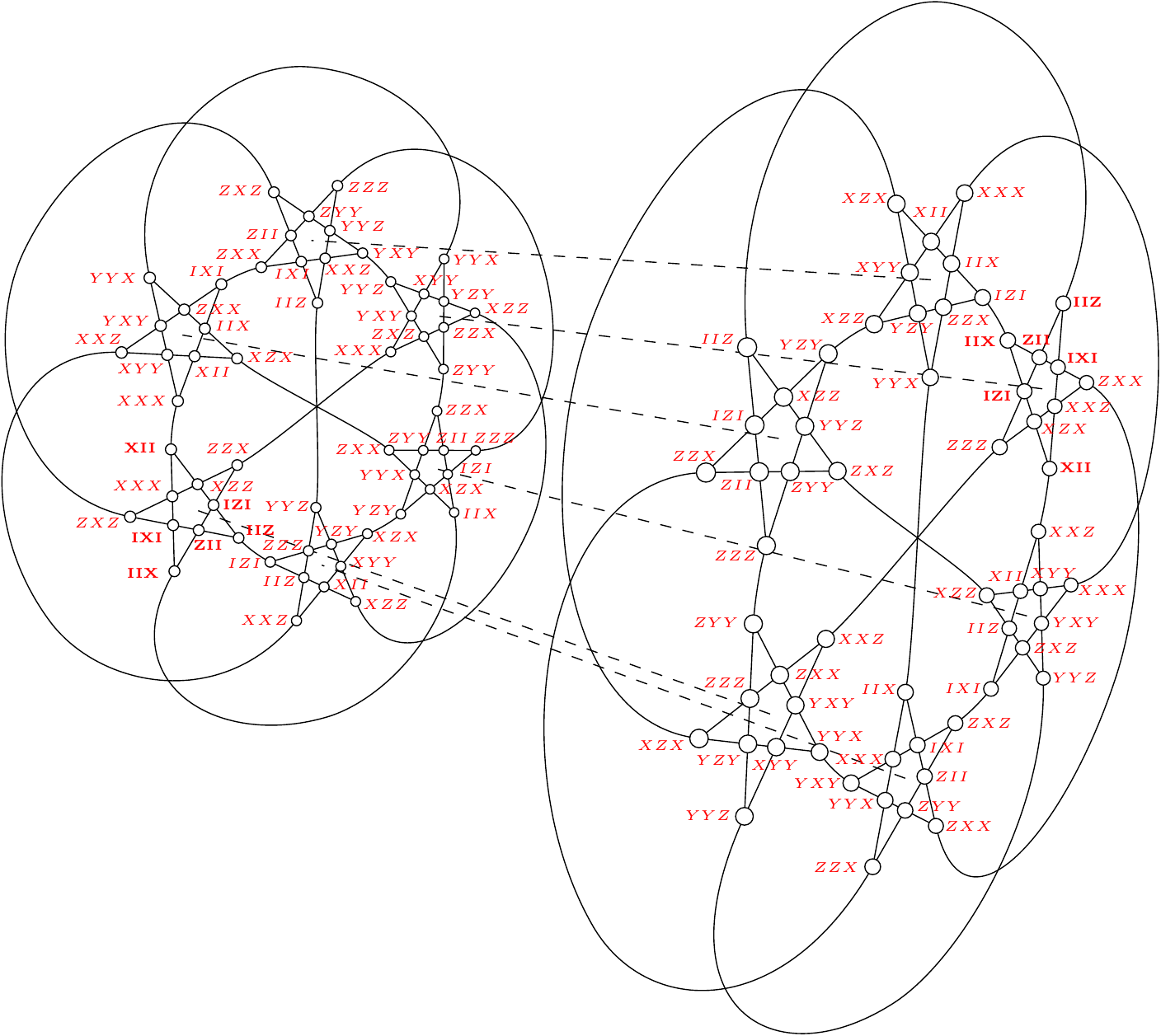}{Double six of pentagrams constructed from the canonical basis\cite{Zsolt} (cf.\ Figure \figref{figure2})}

The product modulo $\mu_4$, or, in terms of the symplectic vector space $V_3$, the sum of the ten observables\footnote{Actually, it suffices to consider only six observables not incident with some line.} of some pentagram does not depend on the choice of pentagram, and it is specific to the double six. Let us denote this observable by $w$. In the case of the example in Figure \figref{figure4}, $w = YYY$. $w$ anticommutes with all 20 observables of the double six, and the antipodal of any pentagram $P$ is the image of $P$ under the transvection $T_w$. The former observation implies that the restriction of $T_w$ to the 20 observables of the double six is the addition of $w$.

Let $\{ e_1, e_2, e_3, f_1, f_2, f_3 \}$ be a symplectic basis the elements of which are not incident with an arbitrary line of an arbitrary pentagram of the double six, and let
\beq p = \sum_{i=1}^3 [Q_0 (e_i) f_i + Q_0 (f_i) e_i ]. \eeq
We assert without proof\footnote{Some of the arguments can be found in \cite{Zsolt}.} that $p$ is also specific to the double six, i.e.\ it does not depend on the choice of pentagram and line, it is symmetric and $q = p+w$ is antisymmetric.
It is easy to check that in our case the double six is characterized by $p=III$ and $q=YYY$.
 Moreover, the pair $p,q$ unambiguously identifies the double six. Since the number of symmetric and antisymmetric three-qubit Pauli operators is 36 and 28 respectively, we get $1008 = 36\cdot 28$ double sixes of pentagrams.
These results will be clear when we relate them to properties of a special Veldkamp line in Section \secref{Veldkamp_lines}.

\section{Double-sixes connected to spreads of the "doily"}
\seclbl{doily_spreads}

The Lagrangian subspaces of $V_3$ were introduced in Section \secref{mathbg}, and we saw that these subspaces correspond to maximal sets of pairwise commuting three-qubit Pauli operators. Suppose we have a Mermin pentagram $P$ and let $l_i$, $i=1,...,5$ denote its lines, i.e.\ the sets of four pairwise orthogonal vectors, realized of course by four pairwise commuting observables. Any three vectors incident with a line $l_i$ are linearly independent, thus $l_i$ spans a Lagrangian subspace $U_i$. If, say, $l_i = \{ a, b, c, d \}$, then $d$, chosen arbitrarily, can be written as
\beq d = a + b + c, \eeq
and the set of nontrivial vectors in $U_i$ is of the form
\beq[back1] \{ a, \ b, \ c, \ a+b, \ b+c, \ a+c, \ a+b+c \}. \eeq
The set of three nontrivial vectors
\beq[back2] L_i = \{ a+b, \ b+c, \ a+c \} \subseteq U_i \setminus l_i \eeq
is a totally isotropic line of $\mathcal W(5,2)$, and each of its elements, being the sum of two vectors \emph{not} orthogonal to $w$ (see Section \secref{double_six}), is orthogonal to $w$.

The sets $L_i$ are mutually disjoint, thus $\{ L_1, L_2, L_3, L_4, L_5 \}$ is a partition of $\mathcal P \equiv \bigcup L_i$. By looking at a different pentagram $P'$ from the same double six $P$ belongs to, one gets the same set $\mathcal P$ of 15 observables. Moreover, if $P'$ is the antipodal of $P$, even the partition of $\mathcal P$ will be the same, i.e.\ one gets
\beq \{ L_1, L_2, L_3, L_4, L_5 \} = \{ L_1', L_2', L_3', L_4', L_5' \}. \eeq
The incidence structure $(\mathcal P, \mathcal L, \in)$, where $\mathcal L$ is the set of every possible set $L_i$ for the respective double six, is a subgeometry of $\mathcal W(5,2)$ well known to finite geometers; it is called the \emph{doily} (Figure \figref{figure5}, left; for more details about the doliy see \cite{Polster}, Section 4.1).

\inputfig{figure5}{The doily and two if its spreads}

By definition, a set of lines that is a partition of $\mathcal P$ is a \emph{spread} of the doily, and there are a total of six spreads. Two of these are shown in Figure \figref{figure5}, middle and right, where the lines contained are highlighted in thick curves, and the remaining four can be depicted as in the rightmost diagram of Figure \figref{figure5} up to a rotation by $n\cdot 72$ degrees, $n = 1,2,3,4$. Thus, each pair $P$, $P'$ of antipodal pentagrams of the double six is related to a spread of the doily. For the case of the double six in Figure \figref{figure4} this relation is illustrated in Figure \figref{figure6}. Figure \figref{figure6} shows only one of the two partitions, and the large circles filled with gray represent points of the doily. The thickness of the curves does not represent anything, it only helps to distinguish the relevant objects.

\inputfig[1.2]{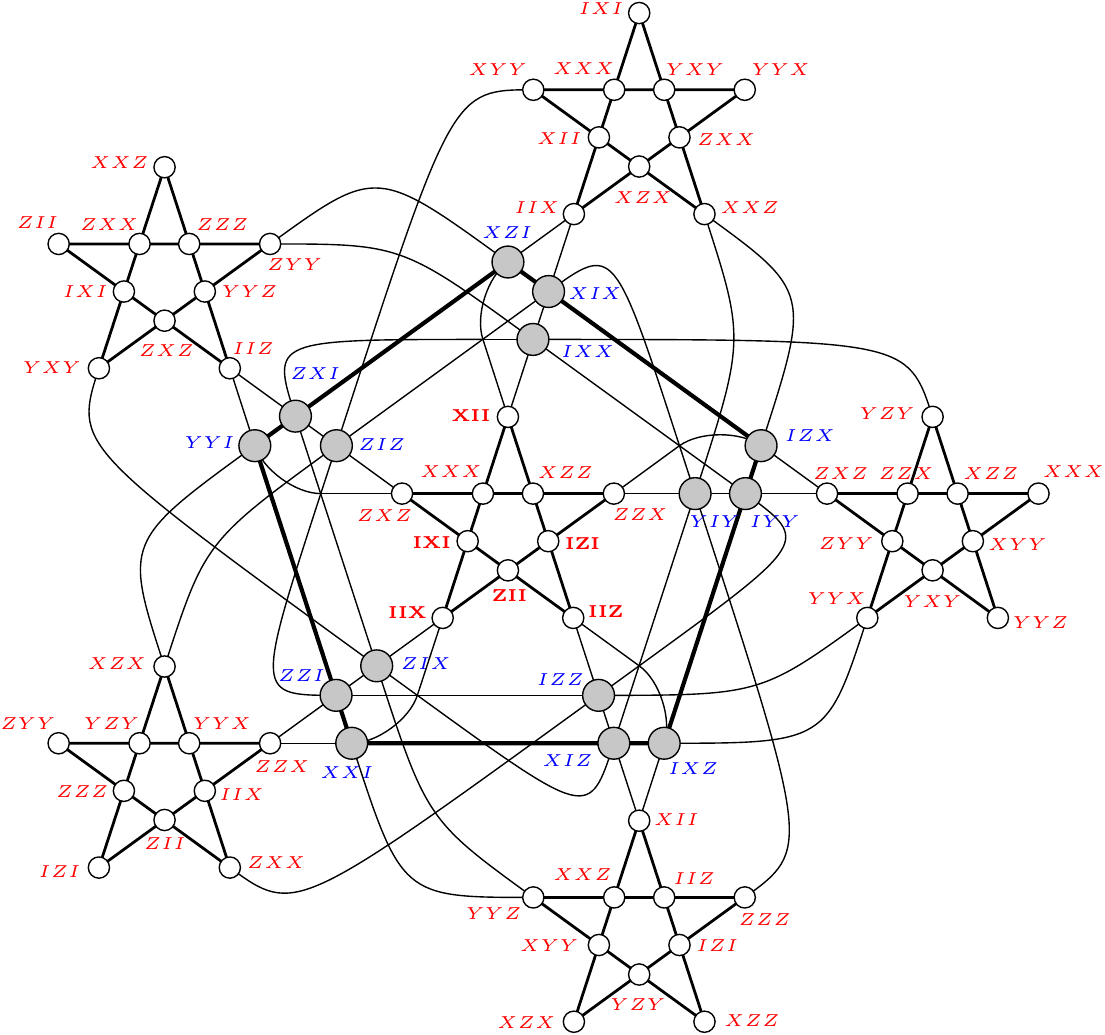}{The doily and one of the two partitions of the double six}

\section{Double-sixes of pentagrams arising from Veldkamp lines}
\seclbl{Veldkamp_lines}

We have seen that $12$ Mermin pentagrams featuring a set of $20$ three-qubit observables can be organized into a $6+6$ configuration dubbed "double-six", which are in turn connected to spreads of a single "doily" a generalized quadrangle $\fnb{GQ}{2,2}$ featuring a set of $15$ further observables. The connection between these incidence structures was effected by the structure of a set of Lagrangian subspaces, representing seven-tuples of mutually commuting sets of observables. Indeed, this $15+20$ split of observables yielded a cut of the relevant Lagrangian subspaces into a $3+4$ structure, where the triples and quadruplets of commuting observables were residing within the doily and the double-six of pentagrams respectively. 

Another crucial observation is that all of these $35=15+20$ observables are {\it symmetric} ones, meaning that the corresponding points in ${\mathcal W}(5,2)$ are lying on the hyperbolic Klein quadric. In other words these $35$ points are lying on a geometric hyperplane of type $H_{III}$.  

We have also seen that in this picture a further observable namely $YYY$ acquired a special status. The transvection $T_{YYY}$ was exchanging the pentagrams taken from one of the sets of six pentagrams with the ones of the other set of six ones. Moreover, $YYY$ was commuting with all of the $15$ observables comprising the doily. Hence the $15$ observables of $H_{III}$ can also be regarded as the ones taken from the $31$ element perp set of another geometric hyperplane namely $C_{YYY}$.

One further piece of evidence comes from \cite{Levfin}. In this work a labelling of the 27 points of the generalized quadrangle $\fnb{GQ}{2,4}$ (alias an elliptic quadric in ${\mathcal W}(5,2)$) in terms of three-qubit observables have been given. According to Figure 3 of \cite{Levfin}, the $27$ points split as $15+6+6$, where the $15$ element set again gives rise to the incidence structure of a doily. The $6+6$ configuration is isomorphic to the double six shown in Figure \figref{figure3}\cite{Levfin}, and the points of this double-six are interchanged via a transvection $T_w$ where $w$ corresponds to the point defining the elliptic quadric, i.e.\ $Q_w(x)=0$. Taking this point to be $w=YYY$ we obtain yet another geometric hyperplane $H_{YYY}$ and one can easily check that for this choice the $15$ points will be precisely the $15$ ones we have already come across with in connection with the hyperplanes $C_{YYY}$ and $H_{III}$.

From these considerations it is clear that what we have described here is just the Veldkamp line $(H_{III}, H_{YYY}, C_{YYY})$ whose core set is the doily labelled with a particular set of symmetric three-qubit observables.

In order to make these considerations very explicit, in the following we give the observables representing the points of the geometric hyperplanes comprising our particular Veldkamp line. First, we list the observables of each hyperplane that \emph{do not belong} to the common core set. The $20$ points taken from the Klein-quadric $H_{III}$ are the symmetric observables
\beq[array20]\begin{matrix}
XXX & XXZ & XZX & ZXX & IIX & IXI & XII & IIZ & IZI & ZII \\
ZZZ & ZZX & ZXZ & XZZ & YYZ & YZY & ZYY & YYX & YXY & XYY
\end{matrix}\eeq
which build up the double-six of pentagrams, where the $10+10$ split refers to the role of the transvection $T_{YYY}$ exchanging the sets. The $12$ points taken from the elliptic quadric $H_{YYY}$ are comprising the antisymmetric observables anticommuting with $YYY$ (see \eqnref{ujkvadforms}). These are
\beq\begin{matrix}
YXI & IYZ & IYX & XIY & ZIY & YZI \\
IZY & YIX & YIZ & ZYI & XYI & IXY
\end{matrix}\eeq
where the $6+6$ split refers to Schl\"afli's double-six configuration with the transvection $T_{YYY}$ exchanging the sets. The $16$ points taken from the perp set $C_{YYY}$ are comprising the antisymmetric observables commuting with $YYY$. These are
\beq\begin{matrix}
YYY & IIY & IYI & YII & ZZY & ZYZ & YZZ & XXY & XYX & YXX \\
XYZ & ZXY & YZX & ZYX & XZY & YXZ &  &  &  &  
\end{matrix}\eeq
And finally, the $15$ points of the common core set are
\beq[array15]\begin{matrix}
XXI & XIX & IXX & YYI & YIY & IYY & ZZI & ZIZ & IZZ & IXZ \\
ZIX & XZI & ZXI & IZX & XIZ &  &  &  &  & 
\end{matrix}\eeq
These observables belong to all of the hyperplanes $H_{III},H_{YYY},C_{YYY}$ and can be used to label the incidence stucture of a doily. Indeed, these observables are symmetric (corresponding to points in $H_{III}$), they are symmetric and commuting with $YYY$ (corresponding to points of $H_{YYY}$), and finally they are commuting with $YYY$ (corresponding to points of $C_{YYY}$). Altogether we have $63= 20+12+16+15$ nontrivial observables (modulo elements of $\mu_4$) yielding the correct number of points of ${\mathcal W}(5,2)$. This structure of our Veldkamp line is highlighted in Figure \figref{figure7}.

\inputfig{figure7}{The Veldkamp line $H_{III}-H_{YYY}-C_{YYY}$}

We already know\cite{VranLev} that the number of Veldkamp lines featuring a hyperbolic and an elliptic quadric is $36\cdot 28=1008$. Each Veldkamp line is associated with a double-six configuration of pentagrams. The structure of a double-six configuration is entirely encoded into the structure of spreads of the core set.

Clearly, no two double-six configurations associated with {\it different} Veldkamp lines contain the same pentagram. Indeed, if we assume the existence of such two pentagrams, they should give rise to the same spreads, hence the same core sets i.e.\ doilies. Hence according to property $V2$ this contradicts our assumption of different Veldkamp lines. In short, the core set is unambiguously characterizes the Veldkamp line hence the total number of pentagrams is $12\cdot 1008=12\,096$. This argument provides a clear geometric understanding of the computer aided result of \cite{Saniga1}. However, apart from our argument producing the right number of pentagrams in a simple manner, it also provides considerable amount of new geometric data on the structure of pentagrams. More importantly for the first time our result establishes a clear physical meaning of the abstract mathematical notion of a Veldkamp line featuring elliptic and hyperbolic quadrics: namely it encodes information on the geometry of all possible Mermin pentagrams that can be built from three-qubit observables.

Let us also show that our argument also produces an explicit algorithm for generating all possible double-sixes from a particular one. Indeed, let us call the Veldkamp line $(H_{III},H_{YYY},C_{YYY})$ the {\it canonical Veldkamp line}: ${\mathcal L}_V^{\ast}$. According to \cite{VranLev} we have

\begin{lem}\label{lem:transitive}
Let $N\in\mathbb{N}+3$, $\mathcal{G}_N=({\mathcal P},{\mathcal L},\in)$ and $a,b,f\in V_{N}$ three
distinct vectors such that $Q_0(a) = Q_0(b)$. Then there exists
an element in $\fnb{Sp}{2N,2}$ fixing $H_{f}$ and swapping $H_{a}$ with $H_{b}$.
\end{lem}

Using this result one can conclude that $\fnb{Sp}{2N,2}$ acts transitively on the set of Veldkamp lines featuring two $H$-type hyperplanes. In our particular case this means that for any two Veldkamp lines from the $1008$ ones one can find an element of $\fnb{Sp}{6,2}$ transforming them to each other. The proof of this Lemma constructs this element of $\fnb{Sp}{6,2}$ by giving the explicit form of the involutions swapping the hyperplanes of the same type.

Since we will use it later let us describe this construction (for a proof see \cite{VranLev}). If $Q_f (a+b) = 1$, the transvection $T_{a+b}$ is an involution, swaps $H_a$ and $H_b$, and fixes $H_f$. If $Q_f (a+b) = 0$, pick an element $p\in C_{a+b} \cap H_a \setminus H_f$, and form $q = p + a + b$. Then $T_p$ and $T_q$ commute, hence $T_p T_q$ is an involution, moreover, $T_p T_q H_a = H_b$, and $H_f$ is again fixed.

As an example needed later we take the following two Veldkamp lines: 
$(H_{III},H_{YYY},C_{YYY})$ and $(H_{YYI}, H_{YYY}, C_{IIY})$.
The first one corresponds to our canonical line ${\mathcal L}_V^{\ast}$. We would like to transform this line to the new one. Now $f=YYY$, $a=III$ and $b=YYI$. Since $Q_f(a+b)=0$ we have to find a $p\in C_{YYI}\cap H_{III}\setminus H_{YYY}$. This means that we have to find a symmetric observable commuting with $YYI$ and anticommuting with $YYY$. As a particular choice let us take $p=XXX$, hence $q=ZZX$. Then (among the many possible involutions) an involution that gives us the job is $T_{ZZX}T_{XXX}$.

Let us group the $20$ observables of \eqnref{array20} to a $1+9+9+1$ split. The meaning of this split will be clarified later. Then the $20$ new observables belonging to the new hyperbolic quadric $H_{YYI}$ are
\beq[trans1]
T_{ZZX}T_{XXX}\begin{pmatrix}
XXX & IZI & ZXX \\
ZII & YYX & XII \\
XZX & IXI & ZZX
\end{pmatrix} = \begin{pmatrix}
XXX & XYX & XZX \\
YXX & YYX & YZX \\
ZXX & ZYX & ZZX\end{pmatrix} \eeq
\beq[trans2]
T_{ZZX}T_{XXX}\begin{pmatrix}
ZZZ & YXY & XZZ \\
XYY & IIZ & ZYY \\
ZXZ & YZY & XXZ
\end{pmatrix} = \begin{pmatrix}
XXZ & XYZ & XZZ \\
YXZ & YYZ & YZZ \\
ZXZ & ZYZ & ZZZ
\end{pmatrix} \eeq
\beq[trans3] T_{ZZX}T_{XXX}IIX=IIX,\qquad T_{ZZX}T_{XXX}YYZ=IIZ. \eeq
For the transformed observables the pattern should be clear: the two groups of $1+9=10$ observables are distinguished by the third qubit observable being either $X$ or $Z$. Apart form the observables $IIX$ and $IIZ$ the remaining $9+9$ ones are organized into $3\times 3$ matrices with the rows and columns are labelled by the first and second qubit observables taking the values: $X,Y,Z$. Since the involutions $T_{ZZX}T_{XXX}$ leave the symplectic form invariant, the commutation properties are left invariant as well. As a result this involution transforms the canonical double-six configuration to a new one featuring a new set of $12$ pentagrams. Repeating these steps one can reach any double-six hence any of the $12\,096$ pentagrams.

\section{The representation theoretic meaning of double-sixes}
\seclbl{representation}

According to the results of the previous section from  the canonical double-six configuration one can generate any such configuration. Hence we are left with the clarification of the geometry of this canonical configuration.

The double six configuration is featuring $20$ observables. According to \eqnref{trans1}-\eqnref{trans3} these observables can be organized into a $1+9+9+1$ structure reminiscent of the representation theoretic content of Freudenthal triple systems displaying a $1+J+J+1$ structure based on cubic Jordan algebras with $J=6,9,15,27$. Such systems have already been explored within the context of quantum entanglement\cite{LevVran,Freud,BHQC,Kru}. The particular values of $J$ correspond to choosing the Jordan algebras as the algebras of Hermitian $3\times 3$ matrices over the reals, complex numbers, quaternions and octonions. The numbers $6,9,15,27$ are the numbers of independent real parameters these matrices have. It can be shown that the $1+J+J+1$ dimensional module constructed in this way gives rise to irreducible representations of dimension $14,20,32,56$ of particular real forms of the groups of type $\fn* C_3$, $\fn* A_5$, $\fn* D_6$, $\fn* E_7$. In particular the $20={6\choose 3}$ dimensional representation of a real form $\fnb{SU}{6}$ of the group $\fn* A_5$ is furnished by the space of real totally antisymmetric tensors (three forms) $P_{\mu\nu\rho}$ in a six dimensional space ${\mathbb R}^6$ ($\mu,\nu,\rho=1,2,\dots 6$). The group action is
\beq
P_{\mu\nu\rho}\mapsto {S_\mu}^{\mu'} {S_\nu}^{\nu'} {S_\rho}^{\rho'} P_{\mu'\nu'\rho'}, \qquad {S_\mu}^{\mu'}\in \fnb{SU}{6}.
\eeq
The $20$ components of $P_{\mu\nu\rho}$ can be organized to two numbers and two $3\times 3$ matrices as follows\cite{LevVran}. Let us adopt the new labelling $\{1,2,3,4,5,6\}\equiv\{1,2,3,\overline{1},\overline{2},\overline{3}\}$. Then the components of $P_{\mu\nu\rho}$ can be split to two $3\times 3$ matrices and two numbers as follows
\beq[szereposztas]
\begin{pmatrix}P_{\overline{1}23}&P_{\overline{1}31}&P_{\overline{1}12}\\P_{\overline{2}23}&
P_{\overline{2}31}&P_{\overline{2}12}\\
P_{\overline{3}23}&P_{\overline{3}31}&P_{\overline{3}12}\end{pmatrix},\qquad
\begin{pmatrix}P_{1\overline{23}}&P_{1\overline{31}}&P_{1\overline{12}}\\P_{2\overline{23}}&
P_{2\overline{31}}&P_{2\overline{12}}\\
P_{3\overline{23}}&P_{3\overline{31}}&P_{3\overline{12}}\end{pmatrix},
\qquad P_{\overline{123}},\qquad P_{123}. \eeq
Now comparing the right hand sides of \eqnref{trans1}-\eqnref{trans3} with \eqnref{szereposztas} one can conjecture that the geometry of the weight diagram of the $20$ dimensional irrep of $\fn* A_5$ and the geometry of the double-sixes might be related. Indeed, the "overline" involution acting like $P_{\overline{1}23}\mapsto P_{1\overline{23}}$ in this picture seems to correspond to the involution $T_{IIY}$. Moreover, in this picture one can take the vector $P_{123}$ as the one labelled by $IIX$ which in turn should somehow correspond to the highest weight vector of the irrep $20$ of $\fnb{SU}{6}$.

Via employing an explicit construction in the following we prove that the geometry of the double-six configuration can indeed be mapped to the geometry of the weight diagram of the $20$ of $\fnb{SU}{6}$. 

The group $\fnb{SU}{6}$ is one of the real forms of the group of type $\fn* A_5$ which has rank five. The five simple roots can be described\cite{Bour} as the vectors living in a five dimensional hyperplane of ${\mathbb R}^6$ with normal vector $n=(111111)$ as follows
\beq[simpleroots]
\alpha_1=e_1-e_2,\qquad\alpha_2=e_2-e_3,\qquad\alpha_3=e_3-e_4,\qquad\alpha_4=e_4-e_5,\qquad\alpha_5=e_5-e_6
\eeq
where the $e_{\mu}, \mu=1,\dots 6$ form the canonical orthonormal basis vectors of ${\mathbb R}^6$. These vectors satisfy $(n,\alpha_{\mu})=0, \mu=1,2,3,4,5$ ($(\cdot,\cdot)$ is the ordinary scalar product in ${\mathbb R}^6$) and are linearly independent hence can be regarded as the basis vectors of the five dimensional hyperplane where the root vectors of $\fn* A_5$ reside. The length of the simple roots is $2$, and they are either orthogonal or having an angle $120$ degrees between them, hence the Cartan matrix\cite{Slansky} and its inverse for $\fn* A_5$ are
\beq[cartanm]
A_{ij}=(\alpha_i,\alpha_j)=\begin{pmatrix}2&-1&0&0&0\\-1&2&-1&0&0\\0&-1&2&-1&0\\0&0&-1&2&-1\\0&0&0&-1&2\end{pmatrix},\qquad
G_{ij}=\frac{1}{6}\begin{pmatrix}5&4&3&2&1\\4&8&6&4&2\\3&6&9&6&3\\2&4&6&8&4\\1&2&3&4&5\end{pmatrix}.
\eeq
An arbitrary weight vector $\Lambda$ living in this five dimensional hyperplane can be expressed as
\beq[dualban] \Lambda=\sum_{i=1}^5\lambda_i\alpha_i \eeq
where the $\lambda_i$ are the components of the weight vector in the {\it dual} basis\cite{Slansky}.
The {\it Dynkin labels} of $\Lambda$ are given by the formula
\beq[Dynkincomp] a_i=\sum_{i=1}^5\lambda_j A_{ji}. \eeq
Clearly the Dynkin labels of the $\alpha_i$ are given by the columns of the Cartan matrix hence
\beq \alpha_1=(2,-1,0,0,0),\qquad \cdots,\qquad \alpha_5=(0,0,0,-1,2). \eeq
This illustrates a theorem of crucial importance that the Dynkin labels of any root or weight are always {\it integers}. In the case of $A,D,E$ type groups the scalar product between any weights can be expressed as\cite{Slansky}
\beq[scalarferde]
(\Lambda,\Lambda')=a_i' G_{ij} a_j = \lambda_i' A_{ij} \lambda_j \eeq
where for simplicity summation was left implicit.

The Dynkin labels of the highest weight $\Lambda^{(0)}$ vector of the $20$ dimensional irrep of $\fn* A_5$ is\cite{Slansky} 
\beq[hw] a_i^{(0)}=(0,0,1,0,0). \eeq
Since only the Dynkin label $a_3^{(0)}$ is different from zero from this highest weight one can subtract the simple root $\alpha_3$. As a result we get a new weight vector in the representation this time $\Lambda^{(1)}$ with Dynkin labels $a_i^{(1)}=(0,1,-1,1,0)$. In the next step one can subtract $\alpha_2$ and $\alpha_4$ etc. As a result of this procedure one obtains the weight diagram for the $20$ of $\fn* A_5$ see Figure \figref{figure8}.

In the following the $20$ weight vectors of this representation will be labelled as follows: $\Lambda^{(0)},\Lambda^{(r)},\Lambda^{(\overline{r})} \Lambda^{(\overline{0})}$, where $r=1,2,\dots,9$. This labelling is reminiscent of the $1+9+9+1$ structure we have already described. Let us denote the Dynkin labels accordingly as $a_i^{(0)}$, $a_i^{(r)}$, $a_i^{(\overline{r})}$, $a_i^{(\overline{0})}$ then we have a list
\beq\begin{aligned}
a_i^{(1)} &= (0,1,-1,1,0), & a_i^{(2)} &= (1,-1,0,1,0), & a_i^{(3)} &= (0,1,0,-1,1), \\
a_i^{(4)} &= (-1,0,0,1,0), & a_i^{(5)} &= (1,-1,1,-1,1), & a_i^{(6)} &= (0,1,0,0,-1), \\
a_i^{(7)} &= (-1,0,1,-1,1),\quad & a_i^{(8)} &= (-1,0,1,0,-1), \quad & a_i^{(9)} &= (1,-1,1,0,-1),
\end{aligned}\eeq
where the remaining $10$ weights are obtained from these ones via flipping the signs. For example $a_i^{(\overline{9})}=(-1,1,-1,0,1)$ and $a_i^{(\overline{0})}=(0,0,-1,0,0)$.

\inputfig{figure8}{Weight diagram of the 20 dimensional irrep of $\mathrm{A}_5$}

As a next step inverting \eqnref{Dynkincomp} we can calculate the components of these weights in the dual basis with the result: $\lambda_i^{(0)},\lambda_i^{(r)},\lambda_i^{(\overline{r})},\lambda_i^{(\overline{0})}$. Using these components in \eqnref{dualban} and expressing the simple roots in the \eqnref{simpleroots} orthonormal basis one obtains the $20$ weights as elements of ${\mathbb R}^6$ as follows
\beq[súlyok6D]
\begin{aligned}
\Lambda^{(0)} &= \frac{1}{2}(1,1,1,-1,-1,-1)\quad, & \Lambda^{(1)} &= \frac{1}{2}(1,1,-1,1,-1,-1), \\
\Lambda^{(2)} &= \frac{1}{2}(1,-1,1,1,-1,-1), & \Lambda^{(3)} &= \frac{1}{2}(1,1,-1,-1,1,-1), \\
\Lambda^{(4)} &= \frac{1}{2}(-1,1,1,1,-1,-1), & \Lambda^{(5)} &= \frac{1}{2}(1,-1,1,-1,1,-1), \\
\Lambda^{(6)} &= \frac{1}{2}(1,1,-1,-1,-1,1), & \Lambda^{(7)} & =\frac{1}{2}(-1,1,1,-1,1,-1), \\
\Lambda^{(8)} &= \frac{1}{2}(-1,1,1,-1,-1,1), & \Lambda^{(9)} &= \frac{1}{2}(1,-1,1,-1,-1,1).
\end{aligned}\eeq
and the remaining $10$ weights $\Lambda^{(\overline{0})},\Lambda^{(\overline{r})}$ are obtained by flipping the signs. Notice that all of these weight vectors are residinging in the five dimensional hyperplane with normal vector $n$ (the sum of the components is zero). The $20={6\choose 3}$ vectors are arising from the possibilities of choosing from the six slots arbitrary three to have minus signs. All of these vectors are lying on the intersection of a five dimensional sphere of radius squared $\frac{3}{2}$ and our five dimensional hyperplane. Notice also that we have 
\beq[antipode]
(\Lambda^{(0)},\Lambda^{(\overline{0})})=(\Lambda^{(r)}, \Lambda^{(\overline{r})})=-\frac{3}{2},\qquad r=1,\dots, 9 \eeq
\noindent
hence by virtue of 
\beq \cos (\Lambda^{(p)}, \Lambda^{(\overline{p})})=\frac{(\Lambda^{(p)}, \Lambda^{(\overline{p})})}{\vert\vert\Lambda^{(p)}\vert\vert  \vert\vert\Lambda^{(\overline{p})}\vert\vert}=-1,\qquad
p=0,1,\dots 9 \eeq
these vectors are {\it antipodal}.

For the remaining pairs of vectors we have
\beq[nonantipode]
\vert(\Lambda^{(p)}, \Lambda^{(q)})\vert =\frac{1}{2},\qquad p\neq \overline{q},\qquad p,q=0,1,\dots 9,\overline{0},\overline{1},\dots,\overline{9}. \eeq
A much more compact labelling of the $20$ weight vectors can be obtained by adopting the labelling of the six component vetors $\{1,2,3,4,5,6\}\equiv\{1,2,3,\overline{1},\overline{2},\overline{3}\}$ as in \eqnref{szereposztas}. By recording only those slots of the six component vectors which have {\it positive} entries one obtains a {\it trivector labelling} of the weight vectors. For example $\Lambda^{(1)}$ has positive entries in slots $124$ i.e.\ $12\overline{1}$ after a permutation we refer to this weight as $\Lambda^{(\overline{1}12)}$. We employed this permutation in order to conform with the trivector labelling convention of \eqnref{szereposztas}. In this way we obtain the dictionary
\beq[szotar1]
\Lambda^{(0)}=\Lambda^{(123)},\qquad
\begin{pmatrix}\Lambda^{(4)}&\Lambda^{(2)}&\Lambda^{(1)}\\\Lambda^{(7)}&\Lambda^{(5)}&\Lambda^{(3)}\\\Lambda^{(8)}&\Lambda^{(9)}&\Lambda^{(6)}\end{pmatrix}=
\begin{pmatrix}\Lambda^{(\overline{1}23)}&\Lambda^{(\overline{1}31)}&\Lambda^{(\overline{1}12)}\\\Lambda^{(\overline{2}23)}&\Lambda^{(\overline{2}31)}&\Lambda^{(\overline{2}12)}\\\Lambda^{(\overline{3}23)}&\Lambda^{(\overline{3}31)}&\Lambda^{(\overline{3}12)}\end{pmatrix}\eeq
and the remaining $10$ weights are obtained by applying the "overline" involution, e.g. $\Lambda^{(\overline{1})}=\Lambda^{(1\overline{12})}$.
Let us now denote by ${\mathcal A},{\mathcal B}$ three element subsets of the set $\{1,2,3,\overline{1},\overline{2},\overline{3}\}$. Then the scalar products of the $20$ weight vectors can be compactly summarized as
\beq
(\Lambda^{(\mathcal A)},\Lambda^{(\mathcal B)})=\begin{cases}-\frac{3}{2}&,\qquad\vert{\mathcal A}\cap{\mathcal B}\vert =0\\
-\frac{1}{2}&,\qquad\vert{\mathcal A}\cap{\mathcal B}\vert =1\\
+\frac{1}{2}&,\qquad\vert{\mathcal A}\cap{\mathcal B}\vert =2\\
+\frac{3}{2}&,\qquad\vert{\mathcal A}\cap{\mathcal B}\vert =3.
\end{cases}
\eeq
Let us now try to map the $20$ weight vectors of \eqnref{szotar1} to the $20$ observables showing up in a double-six of pentagrams arising from the Veldkamp line $(H_{YYI},H_{YYY},C_{IIY})$. For this Veldkamp line we know that the right hand sides of \eqnref{trans1}-\eqnref{trans3} show similar structure to the one of weight vectors. Hence we define the correspondence
\beq[corres1]
\Lambda^{(123)}\leftrightarrow IIX,\qquad
\begin{pmatrix}\Lambda^{(\overline{1}23)}&\Lambda^{(\overline{1}31)}&\Lambda^{(\overline{1}12)}\\\Lambda^{(\overline{2}23)}&\Lambda^{(\overline{2}31)}&\Lambda^{(\overline{2}12)}\\\Lambda^{(\overline{3}23)}&\Lambda^{(\overline{3}31)}&\Lambda^{(\overline{3}12)}\end{pmatrix}
\leftrightarrow
\begin{pmatrix}XXZ&XYZ&XZZ\\YXZ&YYZ&YZZ\\ZXZ&ZYZ&ZZZ\end{pmatrix}
\eeq
and the correspondence for the remaining weights and observables is established by the action of the "overline" involution on the weight side corresponding to the transvection $T_{IIY}$ on the observable side. Hence for example $\Lambda^{(1\overline{23})}\leftrightarrow XXX$ etc. 

Let us consider now the \eqnref{corres1} correspondence
\beq[weightobserv]
\Lambda^{({\mathcal A})}\leftrightarrow {\mathcal O}_{\mathcal A} \eeq
between weight vectors labelled by three element subsets of $\{1,2,3,\overline{1},\overline{2},\overline{3}\}$ and the $20$ observables of the three-qubit Pauli group (modulo elements of $\mu_4$) featuring our double-six. It is easy to check that
\begin{align}
\begin{split}
\vert{\mathcal A}\cap{\mathcal B}\vert ={\rm even} &\leftrightarrow \{{\mathcal O}_{\mathcal A},{\mathcal O}_{\mathcal A}\}=0, \\
\vert{\mathcal A}\cap{\mathcal B}\vert ={\rm odd} &\leftrightarrow [{\mathcal O}_{\mathcal A},{\mathcal O}_{\mathcal A}]=0.
\end{split}
\end{align}
i.e.\ if the intersection size of weight vector labels are even (odd) the corresponding observables are anticommuting (commuting). For example the weight labels $\{\overline{1}23\}$ and $\{\overline{1}31\}$ have intersection size $2$ hence the corresponding  observables $XXZ$ and $XYZ$ are anticommuting, or the weight labels $\{\overline{1}23\}$ and $\{\overline{2}31\}$ have intersection size $1$ hence the observables $XXZ$ and $YYZ$ are commuting. As a last example note that the labels of the highest and lowest weight vectors i.e.\ $\{123\}$ and $\{\overline{123}\}$ are disjoint hence the corresponding observables $IIX$ and $IIZ$ are anticommuting. Clearly every observable is commuting with itself which corresponds to the fact that the intersection size equals $3$.

Notice, that we are merely interested in whether two observables are commuting or anticommuting, which in the finite geometric context translates to the corresponding points beeing collinear or not. This information translates to the incidence structure between weight vectors having scalar product $-\frac{1}{2},\frac{3}{2}$ (commuting), or $\frac{1}{2},-\frac{3}{2}$ (not commuting). Since norm-squared for weight vectors equals $\frac{3}{2}$ the conditions for commuting of observables simplify to the single one that the corresponding weight vectors $\Lambda^{(\mathcal A)}$ and $\Lambda^{(\mathcal B)}$ are incident when the angle between them satisfies
\beq \cos\theta_{{\mathcal A}{\mathcal B}}=-\frac{1}{3} \eeq
which happens when $\vert {\mathcal A}\cap{\mathcal B}\vert = 1$.

Having clarified the meaning of commuting in the weight diagram picture, we should find a condition which corresponds to the one defining the lines of a pentagram. A line of a pentagram is labelled by a pairwise commuting set of four observables with their product producing the trivial observable $III$ (up to a sign). We have $30$ such lines forming the $12$ pentagrams of our double-six. It is easy to check that in terms of four weight vectors this condition translates to
\beq
\Lambda^{({\mathcal A}_1)} + \Lambda^{({\mathcal A}_2)}+\Lambda^{({\mathcal A}_3)} + \Lambda^{({\mathcal A}_4)} = 0, \qquad \vert{\mathcal A}_s\cap {\mathcal A}_t\vert =1,\qquad s,t=1,2,3,4
\eeq
meaning that the sum of four incident weight vectors should produce the zero vector in ${\mathbb R}^6$. In terms of three element subsets this condition means that ${\mathcal A}_1\cup{\mathcal A}_2\cup{\mathcal A}_3\cup{\mathcal A}_4$ should produce the set with {\it double occurrence} for all six labels. Two characteristic examples of that kind are the ones
\begin{align}
\begin{split}
({\mathcal A}_1,{\mathcal A}_2,{\mathcal A}_3,{\mathcal A}_4) &= (\{\overline{123}\},\{\overline{1}23\},\{\overline{2}31\},\{\overline{3}12\}) \\
({\mathcal A}_1,{\mathcal A}_2,{\mathcal A}_3,{\mathcal A}_4) &= (\{\overline{1}23\},\{\overline{2}31\},\{1\overline{31}\},\{2\overline{23}\}).
\end{split}
\end{align}
It is easy to check that in the first case we have $12$ and in the second $18$ possibilities yielding the correct number $30$.

In order to complete the correspondence one can present the weight diagram labelled by observables of our double-six. Using the dictionary of \eqnref{corres1} and Figure \figref{figure8} for the weight diagram one immediately infers that the simple roots are represented by the observables
\beq[srdynkindiag]
(\alpha_1, \alpha_2, \alpha_3, \alpha_4, \alpha_5)\leftrightarrow (IXI,IZI,XXY,ZII,XII). \eeq
These observables can be used to label the Dynkin diagram of $\fn* A_5$, see Figure \figref{figure9}.

\inputfig{figure9}{Dynkin diagram of $\mathrm{A}_5$ labelled with observables}

As one can check the nodes of the Dynkin diagram are connected if the correponding observables are not commuting and not connected if the ones are commuting. To the usual rule of subtracting a certain simple root from a weight, corresponds the rule of applying the transvection $T_{\alpha_i}$ to the observable ${\mathcal O}_{\mathcal A}$  representing the weight $\Lambda^{(\mathcal A)}$. If $[\alpha_i,{\mathcal O}_{\mathcal A}]=0$ then ${\mathcal O}_{\mathcal A}$ is left invariant however, if $\{\alpha_i,{\mathcal O}_{\mathcal A}\}=0$ for some $i$ then ${\mathcal O}_{\mathcal A}$ is transformed to a new observable via multiplying it with the observable which represents $\alpha_i$. For example
\beq
T_{\alpha_3}\Lambda^{(123)}=\Lambda^{(\overline{1}23)}\leftrightarrow T_{XXY}IIX=XXZ
\eeq
due to the fact that $\{XXY,IIX\}=0$ meaning that the corresponding observables are not commuting.

Now as a last step one can display the weight diagram of the $20$ dimensional irrep of $\fnb{SU}{6}$ as labelled by the $20$ observables of our canonical double-six. In order to present this all we have to do is to transform the labels of the Dynkin diagram of Figure \figref{figure9} with the involution $T_{ZZX}T_{XXX}$. With this new labelling of the $\alpha_i$ we have new transvections $T_{\alpha_i}$ acting on the new weights this time represented by the preimage of the $1+9+9+1$ observables showing up in the right hand side of \eqnref{trans1}-\eqnref{trans2}. The resulting structure can be seen in Figure \figref{figure10}. As we have stressed one can regard this as the canonically labelled weight diagram related to or canonical double-six. From this canonical labelling any of the $1008$ labellings related to the full set of double-sixes of pentagrams can be obtained. The only thing we have to do is to apply our familiar method already has been illustrated by our considerations yielding \eqnref{trans1}-\eqnref{trans3}.

\inputfig{figure10}{The canonically labelled weight diagram of the 20 dimensional irrep of $\mathrm{A}_5$}

We have clarified the representation theoretic meaning of our double-six of pentagrams via a rather ad hoc analogy to the structure of a special Freudenthal system. This approach was based on the specific Veldkamp line $(H_{YYI},H_{YYY},C_{IIY})$ and provided an explicit  mapping between weights and observables  as shown in \eqnref{corres1}. In order to get to the same mapping for the {\it canonical} Veldkamp line $(H_{III},H_{YYY},C_{YYY})$ we had to transform back with the involution $T_{ZZX}T_{XXX}$. Can we obtain a more direct approach for the understanding of the mapping of \eqnref{weightobserv} which also gives additional insight into the structure of our class of Veldkamp lines? 

The answer to this question is yes. In order to show this let us consider the following set of generators for a seven dimensional Clifford algebra
\beq[choice]
(\Gamma_1,\Gamma_2,\Gamma_3,\Gamma_4,\Gamma_5,\Gamma_6,\Gamma_7)=(ZYI,YIX,XYI,IXY,YIZ,IZY,YYY) \eeq
satisfying
\beq \{\Gamma_I,\Gamma_J\}=2\delta_{IJ},\qquad I,J=1,2,\dots 7 \eeq
and
\beq[relation5]
i\Gamma_1\Gamma_2\Gamma_3\Gamma_4\Gamma_5\Gamma_6\Gamma_7=III. \eeq
\noindent
Let us then consider the following three sets of operators
\beq
\Gamma_I,\qquad \Gamma_I\Gamma_J,\qquad \Gamma_I\Gamma_J\Gamma_K,\qquad 1\leq I<J<K\leq 7. \eeq
It is easy to check that the first two sets contain $7+21=28$ antisymmetric operators and the third set contains $35$ symmetric ones. Consider now the relations above modulo elements of $\mu_4$. Then the four triangles corresponding to the four subsets of Figure \figref{figure7} giving rise to the explicit list of operators of \eqnref{array20}-\eqnref{array15} can be labelled as
\beq[fontos]
\{\Gamma_{\mu}\Gamma_{\nu}\Gamma_{\rho}\},\qquad \{\Gamma_{\mu},\Gamma_{\mu 7}\},\qquad \{\Gamma_{\mu}\Gamma_{\nu},\Gamma_7\},\qquad \{\Gamma_{\mu}\Gamma_{\nu}\Gamma_7\}\qquad 1\leq \mu<\nu<\rho\leq 6. \eeq
\noindent

One can check that the particular choice of \eqnref{choice} automatically reproduces the set $\{\Gamma_{\mu}\Gamma_{\nu}\Gamma_{\rho}\}$ with our labelling of the $20$ operators of the canonical set in terms of the $3$ element subsets of the set $\{1,2,3,4,5,6\}=\{1,2,3,\overline{1},\overline{2},\overline{3}\}$. Indeed, we have
\beq[corresr2]
\Gamma^{(123)}\leftrightarrow IIX,\qquad
\begin{pmatrix}\Gamma^{(\overline{1}23)}&\Gamma^{(\overline{1}31)}&\Gamma^{(\overline{1}12)}\\\Gamma^{(\overline{2}23)}&\Gamma^{(\overline{2}31)}&\Gamma^{(\overline{2}12)}\\\Gamma^{(\overline{3}23)}&\Gamma^{(\overline{3}31)}&\Gamma^{(\overline{3}12)}\end{pmatrix}
\leftrightarrow
\begin{pmatrix}ZZZ&YXY&XZZ\\XYY&IIZ&ZYY\\ZXZ&YZY&XXZ\end{pmatrix}
\eeq
where we employed the notation $\Gamma^{(\mu\nu\rho)}\equiv\Gamma_{\mu}\Gamma_{\nu}\Gamma_{\rho}$, and as usual the missing entries are obtained by the action of the transvection $T_{YYY}=T_{\Gamma_7}$. Note that by virtue of \eqnref{relation5} this transvection indeed acts as the overline involution since for example
\beq
\Gamma_7\Gamma^{(\overline{1}23)}=\Gamma_7\Gamma_{2}\Gamma_3\Gamma_4\simeq \Gamma_1\Gamma_5\Gamma_6=\Gamma^{(1\overline{23})}
\eeq
where $\simeq$ means equality modulo an element of $\mu_4$. One can also immediately verify that the transvection $T_{\Gamma_7}$ exchanges the two components of the set $\{\Gamma_{\mu},\Gamma_{\mu 7}\}$ (Schl\"afli's double-six) and the $15$ operators provide a labelling for the doily of Figure \figref{figure5}.

One can also realize that the special structure of our canonical Veldkamp line is related to our special realization of our Clifford algebra. Indeed, all of the operators of \eqnref{choice} are {\it antisymmetric} ones. However, since all what is important for us is merely commutation properties, we could have used {\it any} such algebra for producing any other Veldkamp line from our class of 1008 elements. Hence our labelling of \eqnref{fontos} encapsulates the general structure of our particular class of Veldkamp lines. The structure of incidence structures based on Clifford algebras has already been studied\cite{Shaw1,Shaw2}. In particular it is known that the number of such Clifford algebras (modulo permutations) is $288$ and this number splits as $288=280+8$ where $8$ is the number of possibilities for forming a seven dimensional Clifford algebra featuring only antisymmetric operators. The set with $280$ elements is formed by Clifford algebras featuring four symmetric and three antisymmetric operators. An example of this kind is obtained by transforming our Clifford algebra of \eqnref{choice} by the transvection
$T_{ZZX}T_{XXX}$
\beq[choicei7]
(\gamma_1,\gamma_2,\gamma_3,\gamma_4,\gamma_5,\gamma_6,\gamma_7)=(IXX,IYX,IZX,XIZ,YIZ,ZIZ,IIY). \eeq
Of course the operators $\gamma_I$ that can be used to label the operators of the Veldkamp line $(H_{YYI},H_{YYY},C_{YYI})$ give rise to the same patterns as the ones showing up in \eqnref{fontos}.

\section{Conclusions}
\seclbl{conclusions}

In this paper we investigated the geometry of the space of Mermin pentagrams, objects that are well-known for ruling out noncontextual hidden variable theories as alternatives to quantum theory. It is shown that this space of 12\,096 possible pentagrams is organized into 1008 families, with each family containing a "double-six" of pentagrams. We have established an important connection between the structure of the set of these families of Mermin {\it pentagrams} and a particular class of {\it Veldkamp lines} for three qubits. This study provided considerable amount of new geometric data on the structure of pentagrams. More importantly for the first time our result established a clear physical meaning of the abstract mathematical notion of a Veldkamp line for three-qubits featuring elliptic and hyperbolic quadrics. In short: a Veldkamp line of that kind encodes information on the geometry of special arrangements of three-qubit observables, in particular of all possible Mermin pentagrams that can be built from three-qubit observables.

The basic building blocks of the space of Mermin pentagrams are double-sixes. We have shown that they are inherently connected to the structure of the weight diagram for the $20$ dimensional irrep of $\fnb{SU}{6}$. Double-sixes are comprising a substructure within each Veldkamp line featuring $20$ observables. We demonstrated that due to a transitive action of the symplectic group on the $1008$ element set of Veldkamp lines it is enough to study merely one of these double-sixes (a one we called {\it canonical}), the rest of them can be generated in a straightforward manner.

There are many interesting ideas that follow from our considerations. Let us mention just two, the first of them is related to physics and the second to mathematics.

The first of them is the fact that our approach relates the space of Mermin pentagrams to a subset of the space of Mermin squares both built from three-qubit observables. Indeed, let us consider the $9$ observables on the right hand side of \eqnref{trans1} taken together with the observable $IIX$. This observable is commuting with all of the $9$ observables. These $10$ observables form a characteristic substructure from the $1+9+9+1$ structure we have studied. Moreover, $IIX$ taken together with any three observables taken from any of the three diagonals or anti-diagonals of the matrix from the right hand side of \eqnref{trans2} forms six sets of quadruplets with pairwise commuting entries. They are six copies from the $30$ possible pentagram lines taken from a double-six of pentagrams. Multiplying the observables on the right hand side of \eqnref{trans1} with $IIX$ and rearranging the observables as follows one gets
\beq
\begin{pmatrix}
XXI & XYI & XZI \\
YXI & YYI & YZI \\
ZXI & ZYI & ZZI
\end{pmatrix}\mapsto \begin{pmatrix}
XXI & YYI & ZZI \\
YZI & ZXI & XYI \\
ZYI & XZI & YXI
\end{pmatrix}.
\eeq
Let us call for the arrangement on the right hand side the rows and columns of the matrix as the {\it lines} of a grid with the operators being its {\it points}. Then one can check that the grid labelled by three-qubit observables has the properties (P1)-(P3) of Section \secref{double_six} provided this time (P1) means: three pairwise commuting operators that multiply to $\pm III$. The object we have obtained is called: a Mermin-square. One can transform this object back by the transvection $T_{ZZX}T_{XXX}$ with the result
\beq
\begin{pmatrix}XXI&YYI&ZZI\\XIX&XZI&IZX\\IXX&ZXI&ZIX\\\end{pmatrix}
\eeq
being another Mermin-square. In this way for each Veldkamp line one obtains a Mermin-square $1008$ in total. However, in our construction the observable $IIX$ (the highest weight) was special. But nothing prevents us from repeating the same construction by elevating {\it any} of the $20$ observables to the status of a highest weight. Hence the total number of Mermin squares obtained in this way seems to be $20\,160$. However, since the highest and lowest weight yields the same Mermin-square the correct number is $10\,080$. Interestingly this number is just the half of the total number of Mermin-squares that can be built from three-qubit observables\cite{Planatpriv}.

One can understand this result in yet another way. Let us recall the relevant results of Section \secref{doily_spreads}, in particular \eqnref{back1}-\eqnref{back2}. Mermin pentagrams per Veldkamp line are built from the $20$ observables forming $30$ lines comprising {\it quadruplets} of observables residing in the double-six part of the Veldkamp line. Mermin-squares per Veldkamp line on the other hand are built from the $15$ observables forming the $15$ lines comprising {\it triples} of observables residing in the doily (the core set) part of the Veldkamp line. It is well-known\cite{San2} that a particularily labelled doily is containing $10$ Mermin-squares (grids) as geometric hyperplanes. Since we have $1008$ such doilies as core sets the naive count gives $10\,080$ Mermin squares. The precise nature of this Mermin-square vs. Mermin-pentagram correspondence will be addressed in future work. 

The second of the ideas that follows from our work is representation theoretical. From our investigations it is clear that the notion of a particular Veldkamp line for three-qubits contains valuable representation theoretic information in a finite geometric manner. Apart from the structure of the double-six of pentagrams part which encodes the weight diagram of the $20$ of $\fn* A_5$, it is clear that the Klein-quadric (ie. the $H_{III}$) part featuring $35$ observables should be related to the $35$ of $\fn* A_6$. Indeed,  due to ${7\choose 3}=35$ the relevant representation in this case is the one on trivectors in a {\it seven} dimensional space. On the other hand, as already mentioned in the Introduction the  elliptic quadric part of our canonical Veldkamp line (i.e.\ the $H_{YYY}$ part) is encoding information on the weight diagram of the $27$ of $\fn* E_6$. Hence by a convenient labelling for the $\fn* E_6$ Dynkin diagram one should be able to reproduce a labelling of the weight diagram of the $27$ dimensional irrep in a manner similar to Figure 
\figref{figure9}.

This connection between representation theory and finite geometry should of course survive for other Veldkamp lines and other multiqubit observables. In particular we conjecture that the $N=2,3,4$ cases should produce some of the low dimensional representations of the simply laced groups (the ones of ADE-type) in a finite geometric setting. We are planning to explore these fascinating ideas in a future work. 

Finally, let us note that our findings can also pave the way for further studies for uncovering the connection found between noncontextual configurations and weight diagrams of certain representations of Lie-groups. We even conjecture that by studying other Veldkamp lines of the Veldkamp space for $N$-qubits makes it possible to find new contextual configurations. This observation might be a starting point to study a certain subset of such configurations in a systematic manner. Our mathematical framework based on the notion of Veldkamp space encoding information on contextual configurations via representation theory might establish a unified picture. Our hope is that our Veldkamp space based approach might even serve as a guiding principle for organizing some of the scattered results relating group representations with contextuality.

\section{Acknowledgements}
We would like to express our gratitude to Istv\'an L\'aszl\'o for cross-checking
our findings concerning the geometry of the double-six of pentagrams using his computer program. Zs.\ Sz.\ is grateful to P\'eter L\'evay for his patience and kind support over the years.


\begin{thebibliography}{}

\bibitem{KS} E. P. Specker, Dialectica {\bf 14}, 239 (1960); S. Kochen, E. P. Specker,
J. Math. Mech. {\bf 17}, 59 (1967).
\bibitem{Bell} J. S. Bell, Rev. Mod. Phys. {\bf 38}, 447 (1966);
Reprinted in J. S. Bell, {\it Speakable and Unspeakable in Quantum Mechanics}, Cambridge University Press, Cambridge, UK, 1987.
\bibitem{Peres} A. Peres, J. Phys. A {\bf 24}, L175 (1991).
\bibitem{Peresbook} A. Peres, {\it Quantum Theory: Concepts and Methods}, Kluwer Academic Publishers, (2002).
\bibitem{Mermin1990} N. D. Mermin, Phys. Rev. Lett. {\bf 65}, 3373 (1990).
\bibitem{Mermin1993} N. D. Mermin, Rev. Mod. Phys. {\bf 65}, 803 (1993).
\bibitem{Horne} D. M. Greenberger, M. A. Horne and A. Zeilinger in {\it Bell's Theorem, Quantum Theory and Conceptions of the universe}, edited by M. Kafatos (Kluwer, Dordrecht, 1989);
D. M. Greenberger, M. A. Horne, A. Shimony and A. Zeilinger: Am. J. Phys. {\bf 58}, 1131 (1990).
\bibitem{Waegell} M. Waegell and P. K. Aravind, Phys. Lett. A{\bf 377}, 546 (2013);
J. Phys. A: Math. Theor. 45, 405301 (2012); Phys. Rev. A{\bf 88}, 012102 (2013).
\bibitem{Waegell3qbit} M. Waegell and P. K. Aravind, J. Phys. A: Math. Theor. {\bf 45} 405301 (2012). \url{http://arxiv.org/abs/1205.5015}
\bibitem{Saniga1} M. Planat, M. Saniga and F. Holweck, Quantum Information Processing {\bf 12}, 2535 (2013). \url{https://arxiv.org/abs/1212.2729}
\bibitem{Levay1} P. L\'evay, M. Planat and M. Saniga, Journal of High Energy Physics {\bf 9}, 037 (2013). \url{http://arxiv.org/abs/1305.5689}
\bibitem{Planat} M. Planat, Information {\bf 5} 209 (2014).
\bibitem{SLev} M. Saniga and P. L\'evay, Europhysics Letters {\bf 97} 50005 (2012).
\bibitem{Levfin} P. L\'evay, M. Saniga, P. Vrana, Phys. Rev. D{\bf78} 124022, (2008). \url{http://arxiv.org/abs/0808.3849}
\bibitem{Levfin2}
P. L\'evay, M. Saniga, P. Vrana, P. Pracna,  Phys. Rev. D{\bf 79} 084036, (2009).  \url{http://arxiv.org/abs/0903.0541}
\bibitem{BHQC} L. Borsten, M. J. Duff and P. L\'evay, Classical and Quantum Gravity 29 (22), 224008 (2012).
\bibitem{San1} M. Saniga, M. Planat, Quantum Information and Computation {\bf 8}, 127 (2008), Advanced Studies in Theoretical Physics {\bf 1}, 1 (2007). 
\bibitem{Buek}Buekenhout F., Cohen A.M., Diagram geometry, Springer (2009)
\bibitem{Nielsen} M. A. Nielsen and I. L. Chuang, {\it Quantum Computation and Quantum Information}, Cambridege University Press 2000.
\bibitem{Gottesman} D. Gottesman, Phys. Rev. {\bf A54}, 1862 (1996), 
D. Gottesman, Phys. Rev. {\bf A57} 127 (1998).
\bibitem{Sloane} A. R. Calderbank, E. M. Rains, P. W. Shor and N. J. A. Sloane,
Phys. Rev. Lett. {\bf 78} 405 (1997).
\bibitem{Shult} E. Shult, Bull. Belg. Math. Soc. {\bf 4}, 299 (1997).
\bibitem{Veld}  E. Shult, Points and Lines: Characterizing the Classical Geometries, Chapter 4.1., 
Springer-Verlag Berlin Heidelberg (2011).
\bibitem{San2} M. Saniga and M. Planat, P. Pracna, SIGMA {\bf 3}, 75 (2007). \url{http://arxiv.org/abs/0704.0495v3}
\bibitem{VranLev} P. Vrana, P, L\'evay, J. PhysA: Math. Theor. A{\bf 43}, 125303 (2010). \url{http://arxiv.org/abs/0906.3655v3}.
\bibitem{Bour} N. Bourbaki, {\it Elements of Mathematics, Lie Groups and Lie Algebras}, Chapters 4-6, Masson, Springer-Verlag Berlin Heidelberg (2002), see on page 243.
\bibitem{SpGen} O. T. O'Meara, {\it Symplectic groups}, Chapter 2.1, American Mathematical Society, USA (1978)
Yaim Cooper. {\it Generators of the Symplectic Group}. 2005. \url{http://www-math.mit.edu/~dav/sympgen.pdf}.
\bibitem{Geemen} B. L. Cherchiai, B. van Geemen, J. Math. Phys.  {\bf 51}, 122203 (2010), \url{http://arxiv.org/abs/1003.4255}
\bibitem{Payne}  S. E. Payne and J. A. Thas, {\it Finite Generalized Quadrangles}, Pitman Boston-London-Melbourne (1984).

\bibitem{Duff} M. J. Duff, S. Ferrara,  Phys. Rev. D{\bf 76} 124023 (2007).
\bibitem{HS} F. Holweck and M. Saniga, {\it Contextuality with a Small number of Observables,} 
\url{http://arXiv.org/abs/1607.07567v1}
\bibitem{Zsolt} Zsolt Szab\'o, {\it The finite Geometric Aspects of Quantum Contextuality}, BSc Thesis (in Hungarian),
Institute of Physics, Budapest University of Technology and Economics (2016).
\bibitem{Polster} B. Polster, {\it A Geometric Picture Book}, Springer-Verlag New York (1998).
\bibitem{LevVran} P. L\'evay and P. Vrana, Phys. Rev. A{\bf 78}, 022329 (2008). \url{http://arxiv.org/abs/0806.4076v1}
\bibitem{Freud}P. Vrana and P. L\'evay, J. Phys. A: Math. Theor. 42, 285303 (2009). \url{http://arxiv.org/abs/0902.2269v2}
\bibitem{Kru} S. Krutelevich, Journal of Algebra, {\bf 314}, 924 (2007).
\bibitem{Slansky} R. Slansky, Physics Reports {\bf 79}, 1-128 (1981).
\bibitem{Shaw1} R. Shaw, {\it Finite geometry, Dirac groups and the table of real Clifford algebras} pp. 59-99, in R. Ablamowicz and P. Lounesto, eds., Clifford Algebras and Spinor Structures (Kluwer Acas. Pubs., Dordrecht, 1995).
\bibitem{Shaw2} R. Shaw, J. Phys. A: Math. Gen. {\bf 21} 7-16 (1988).
\bibitem{Planatpriv} M. Planat, private communication (2016).
\end{thebibliography}
\end{document}